\documentclass[fleqn,aps,prl,nofootinbib,twocolumn,letterpaper,superscriptaddress]{revtex4-2}
\usepackage{dcolumn}
\usepackage{color}
\usepackage{graphicx}
\usepackage[colorlinks=true]{hyperref}
\usepackage{multirow}
\usepackage{float}
\usepackage{lineno}
\usepackage{hyperref}
\usepackage{booktabs}
\usepackage{amsmath,amsthm,amsfonts,amssymb,bm}

\begin{document}
\def\shKeyLab{School of Physics and Astronomy, Shanghai Jiao Tong University, Key Laboratory for Particle Astrophysics and Cosmology (MoE), Shanghai Key Laboratory for Particle Physics and Cosmology, Shanghai 200240, China}
\def\scKeyLab{Jinping Deep Underground Frontier Science and Dark Matter Key Laboratory of Sichuan Province}
\def\BUAA{School of Physics, Beihang University, Beijing 102206, China}
\def\BUAACenter{Peng Huanwu Collaborative Center for Research and Education, Beihang University, Beijing 100191, China}
\def\BUAALab{Beijing Key Laboratory of Advanced Nuclear Materials and Physics, Beihang University, Beijing, 102206, China}
\def\SCNT{Southern Center for Nuclear-Science Theory (SCNT), Institute of Modern Physics, Chinese Academy of Sciences, Huizhou 516000, China}
\def\USTClab{State Key Laboratory of Particle Detection and Electronics, University of Science and Technology of China, Hefei 230026, China}
\def\USTCdep{Department of Modern Physics, University of Science and Technology of China, Hefei 230026, China}
\def\BUAALab{International Research Center for Nuclei and Particles in the Cosmos \& Beijing Key Laboratory of Advanced Nuclear Materials and Physics, Beihang University, Beijing 100191, China}
\def\pku{School of Physics, Peking University, Beijing 100871, China}
\def\YaLongSD{Yalong River Hydropower Development Company, Ltd., 288 Shuanglin Road, Chengdu 610051, China}
\def\IAP{Shanghai Institute of Applied Physics, Chinese Academy of Sciences, 201800 Shanghai, China}
\def\CHEPpku{Center for High Energy Physics, Peking University, Beijing 100871, China}
\def\SDUdep{Research Center for Particle Science and Technology, Institute of Frontier and Interdisciplinary Science, Shandong University, Qingdao 266237, Shandong, China}
\def\SDUlab{Key Laboratory of Particle Physics and Particle Irradiation of Ministry of Education, Shandong University, Qingdao 266237, Shandong, China}
\def\UMD{Department of Physics, University of Maryland, College Park, Maryland 20742, USA}
\def\TDLee{New Cornerstone Science Laboratory, Tsung-Dao Lee Institute, Shanghai Jiao Tong University, Shanghai 201210, China}
\def\MESJTU{School of Mechanical Engineering, Shanghai Jiao Tong University, Shanghai 200240, China}
\def\SYU{School of Physics, Sun Yat-Sen University, Guangzhou 510275, China}
\def\SYUSFI{Sino-French Institute of Nuclear Engineering and Technology, Sun Yat-Sen University, Zhuhai, 519082, China}
\def\NKU{School of Physics, Nankai University, Tianjin 300071, China}
\def\YTU{Department of Physics, Yantai University, Yantai 264005, China}
\def\FDU{Key Laboratory of Nuclear Physics and Ion-beam Application (MOE), Institute of Modern Physics, Fudan University, Shanghai 200433, China}
\def\USST{School of Medical Instrument and Food Engineering, University of Shanghai for Science and Technology, Shanghai 200093, China}
\def\SJTUSC{Shanghai Jiao Tong University Sichuan Research Institute, Chengdu 610213, China}
\def\SPEIT{SJTU Paris Elite Institute of Technology, Shanghai Jiao Tong University, Shanghai, 200240, China}
\def\NNU{School of Physics and Technology, Nanjing Normal University, Nanjing 210023, China}
\def\SYSUzhuhai{School of Physics and Astronomy, Sun Yat-Sen University, Zhuhai 519082, China}
\def\CDUT{College of Nuclear Technology and Automation Engineering, Chengdu University of Technology, Chengdu 610059, China}

\affiliation{\TDLee}
\author{Xinning Zeng}\affiliation{\shKeyLab}
\author{Zihao Bo}\affiliation{\shKeyLab}
\author{Wei Chen}\affiliation{\shKeyLab}
\author{Xun Chen}\affiliation{\TDLee}\affiliation{\shKeyLab}\affiliation{\SJTUSC}\affiliation{\scKeyLab}
\author{Yunhua Chen}\affiliation{\YaLongSD}\affiliation{\scKeyLab}
\author{Zhaokan Cheng}\affiliation{\SYUSFI}
\author{Xiangyi Cui}\affiliation{\TDLee}
\author{Yingjie Fan}\affiliation{\YTU}
\author{Deqing Fang}\affiliation{\FDU}
\author{Zhixing Gao}\affiliation{\shKeyLab}
% ...
\collaboration{PandaX Collaboration}
\noaffiliation

\title{Exploring New Physics with PandaX-4T Low Energy Electronic Recoil Data}
\date{\today}

\begin{abstract}
 New particles beyond the standard model of particle physics, such as axions, can be effectively searched through their interactions with electrons.
 We use the large liquid xenon detector PandaX-4T to search for novel electronic recoil signals induced by solar axions, neutrinos with anomalous magnetic moment, axionlike particles, dark photons, and light fermionic dark matter.
 A detailed background model is established using the latest datasets with 1.54 $\rm ton \cdot yr$ exposure.
 No significant excess above the background has been observed, and we have obtained competitive constraints for axion couplings, neutrino magnetic moment, and fermionic dark matter interactions.   
\end{abstract}
\maketitle

Large liquid xenon (LXe) detectors located deep underground have unique advantages in searching for rare signals from dark matter (DM) and neutrino-related physics~\cite{P4_WIMP_Run0Run1, P4_B8_Run0Run1, XENONnT_WIMP, XENONnT_B8, LZ_WIMP}.
In addition to the traditional DM-induced nuclear recoil (NR) events, particles from many new physics models can produce novel electronic recoil signals (NERS), which have attracted increasing attention recently~\cite{P2_axion, FDM_ZhangDan, XENONnT_axion, LZ_SA}.
Axions~\cite{PecceiQuinn, Weinberg, Wilczek} can be produced in the Sun via the ABC effect (atomic recombination and deexcitation, bremsstrahlung, and Compton scattering processes)~\cite{FluxFromAE, AxioElectricEffect}, the Primakoff conversion $\gamma + Ze \rightarrow Ze + a$~\cite{Primakoff}, and the nuclear magnetic transitions~\cite{Fe57Proposal}, related to the axion-electron coupling $g_{ae}$, axion-photon coupling $g_{a\gamma\gamma}$, and effective axion-nucleon coupling $g_{aN}^{\text{eff}}$ $(\equiv -1.19g^0_{aN} + g^3_{aN})$, respectively. 
The corresponding effective Lagrangian is given by~\cite{EDELWEISS_III_axion}:
\begin{equation}
 L = ig_{ae} \bar{e} \gamma_5 e a
 -\frac{1}{4}g_{a\gamma\gamma} F\tilde{F}a 
 + i \bar{N} \gamma_5 (g^0_{aN}+g^3_{aN}\tau_3) N a,
\end{equation}
where $a$ is the axion field, $e$ the electron, $F$ the photon, and $N$ the nucleon isospin doublet $(p, n)$.
$g^0_{aN}$ and $g^3_{aN}$ are model-dependent isoscalar and isovector axion-nucleon couplings and $\tau_3$ is the Pauli matrix.
The Sun also produces a vast number of neutrinos, which may carry anomalous magnetic moment $\mu_\nu$ due to the Majorana nature of neutrino or exotic new physics beyond the standard model (SM)~\cite{Giunti:2014ixa}.
With their kinetic energy governed by the Sun's temperature, these particles can produce electronic recoil (ER) events with typical energy from keV to tens of keV. 
Within this energy range, alternative light dark matter models such as axionlike particles (ALPs), dark photons~\cite{DarkPhoton}, or light fermionic DM converting into active neutrinos (DM-$\nu$)~\cite{Dror_2020,FDM_Dror} can also manifest themselves with characteristic ER spectra. 
ALPs interact with SM particles similarly to axions, while dark photons interact through kinetic mixing $\kappa$ with SM photons.
The vector and axial-vector interactions among the dimension-six effective operators of DM-$\nu$ conversion can be probed experimentally. Considering the atomic effect, the differential cross section with respect to the ionized electron energy $T_r$ is
\begin{equation}
\begin{aligned}
    \frac{d \langle \sigma_{\chi e} v_{\chi} \rangle }{d T_r}  &= \sum_{n,l} (4l + 2) \frac{1}{T_r} \frac{m_{\chi} - T_r-E_{nl}}{16 \pi m_e^2 m_{\chi}} \\
 &\times |\mathcal{M}(\boldsymbol{q})|^2 K_{nl}(T_r, |\boldsymbol{q}|),
\end{aligned}
\end{equation}
where $n$ and $l$ are the principal quantum number and angular momentum.
|$\boldsymbol{q}|$ is the energy of outgoing neutrino and $E_{nl}$ is the energy of the accompanying x-ray, which has the same energy as the binding energy of state $| nl \rangle$.
$|\mathcal{M}(\boldsymbol{q})|^2$ is the particle scattering amplitude and $K_{nl}(T_r, |\boldsymbol{q}|)$
is the atomic $K$ factor~\cite{geRevisitingFermionicDark2022}.
The masses of electron and DM particle are denoted as $m_e$ and $m_\chi$, respectively.

\begin{figure}[tb]
    \centering
    \includegraphics[width=\columnwidth]{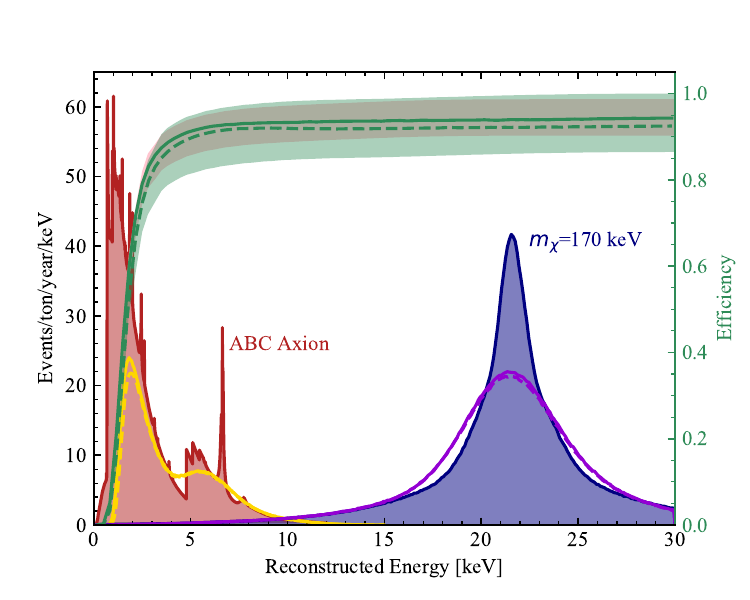}
    \caption{Theoretical (shaded) and expected energy spectra (solid for Run0, dashed for Run1) for ABC axion with a coupling $g_{ae}$=3.1$\times 10^{-12}$ (red shaded for theoretical, yellow unshaded for expected) and  DM-$\nu$ conversion with $m_{\chi}$ = 170 keV and $\sigma_e v_{\chi} = 9.0\times10^{-49}$ cm$^2$ (blue shaded for theoretical, purple unshaded for expected).
 The solid and dashed green lines depict the signal detection efficiencies for Run0 and Run1, including reconstruction, quality cuts, and ROI, with their respective pink and green bands indicating the uncertainties.}
    \label{fig:eff_qc}
\end{figure}

This Letter reports an updated NERS search using the low-energy ER data below 30 keV from PandaX-4T, an LXe detector operating in the China Jinping Underground Laboratory (CJPL), with a total exposure of 1.54 ton$\cdot$yr. 
In different NERS analyses, the same method of searching for spectral features on top of the expected background is used. 
We leave detailed production and detection cross sections and expected NERS spectra in the Supplemental Material~\cite{supp} for brevity. 
To set the stage, the signals of ABC solar axions and the DM-$\nu$ conversion of $m_\chi = 170\,$keV/$c^2$ are shown in Fig.~\ref{fig:eff_qc}, with spectra at the low and high end of the region of interest (ROI). 
The keys to this analysis lie in the understanding of the detector responses, as well as robust prior knowledge of the background from data outside the ROI.

\nocite{CAST_2013}
\nocite{pe_xsec}
\nocite{IP_3channels}
\nocite{Buch_BraggScat}
\nocite{IPE_Theory_Ge}
\nocite{RHFFormFactor} 
\nocite{RHF_Xray}
\nocite{IPE_RHFFormFactor}
\nocite{Wong:2004sp}
\nocite{Beacom:1999prl}
\nocite{NuMagAtomicEff1}
\nocite{NuMagAtomicEff2}
\nocite{NuMagAtomicEff3}
\nocite{ALP_Pospelov}
\nocite{AN2015331}
\nocite{geActiveSterileConversion}
\nocite{Kfactor}

\begin{table*}[tbp]
    \centering
    \caption{Operating conditions of Run0 and Run1. Duration represents the calendar days, and live time refers to the data-taking time after removing low-quality data periods. $\langle\tau_e\rangle$ is the average electron lifetime of each dataset. $\rm V_{cathode}$ and $\rm V_{gate}$ are voltages of the cathode and gate. $g_1$ and $g_{2_b}$ values  are determined by likelihood fits to low energy calibration data~\cite{P4SignalModel}.}
    \resizebox{\linewidth}{!}{
    \renewcommand\arraystretch{1.1}
    \setlength{\tabcolsep}{2mm}{
    \begin{tabular}{c|ccccc|cccccc}
    \toprule[1.5pt]
 Run & \multicolumn{5}{c}{Run0} \vline & \multicolumn{6}{c}{Run1}\\
    \midrule[1pt]
 Set      & 1 & 2 & 3 & 4 & 5 & 1 & 2 & 3 & 4 & 5 & 6\\
    \midrule[1pt]
 Duration (days) & 1.96 & 13.54 & 5.53 & 37.22 & 36.61 & 9.41 & 10.50 & 6.18 & 14.00 & 38.44 & 85.08 \\
 Live Time (days) & 1.82 & 12.30 & 2.14 & 33.01 & 34.29 & 8.07 & 8.53 & 5.56 & 12.19 & 34.45 & 77.86 \\
    $\langle\tau_{e}\rangle(\mu s)$ & 807.2 & 958.1 & 812.5 & 1110.4 & 1275.4 & 631.2 & 642.3 & 975.7 & 831.5 & 1171.7 & 1719.6 \\
    $\rm V_{cathode}$(-kV) & 20 & 18.6 & 18 & 16 & 16 & \multicolumn{6}{c}{16}\\
    $\rm V_{gate}$(-kV) & 4.9 & 4.9 & 5 & 5 & 5 & \multicolumn{6}{c}{6} \\
    \midrule[1pt]
 $g_1(\%)$  & \multicolumn{5}{c}{$10.0 \pm 0.5$} \vline & \multicolumn{6}{c}{$9.1 \pm 0.4$} \\
    $g_{2_b}$(PE/${e^-}$) & \multicolumn{5}{c}{$4.1 \pm 0.4$} \vline & \multicolumn{6}{c}{$5.0 \pm 0.5$} \\
    \bottomrule[1.5pt]
    \end{tabular}
 }
 }
    
    \label{tab:basic_conf}
\end{table*}

The PandaX-4T utilizes a time projection chamber (TPC), with 3.7 ton of LXe in the sensitive volume, to measure the precise position, timing, and energy of each event~\cite{P4_TDR}.
Energy release in xenon produces prompt scintillation photon signals ($S1$) and ionized electrons.
Electrons drift upward under the influence of a downward electric field in the sensitive volume, and the delayed electroluminescence signals ($S2$) are generated in the gaseous xenon region. 
Photons from both signals are collected by arrays of Hamamatsu R11410-23 three-inch photomultiplier tubes (PMTs) at the top and bottom of the TPC.
Combining the photon pattern on the PMT arrays and the time delay between $S1$ and $S2$, the three-dimensional vertex of a given event can be deduced. 

The electron-equivalent recoil energy $E_{ee}$ is reconstructed as 
\begin{equation}
 E_{ee} = W_q \times (\frac{Q^c_{S1}}{g_1} + \frac{Q^c_{S2_{b}}}{g_{2_{b}}}),
 \label{equ:Eee}
\end{equation}
in which $Q^c_{S1}$ ($Q^c_{S2_{b}}$) is the corrected charge of $S1$ ($S2$ collected by the bottom PMTs), and $W_q = 13.7$ eV is the mean energy to produce a single photon or ionized electron in liquid xenon.
Using $S2_{b}$ avoids the influence of potential saturation and dead channel effects from the top PMTs.
$g_1$ is the photon detection efficiency for $S1$. 
$g_{2_b}$ is the product of electron extraction efficiency and the single-electron gain of $S2_b$.
Both parameters are determined via calibrations~\cite{P4SignalModel, P4_WIMP_Run0Run1}.

For this analysis, we utilize the data from Run0 (Nov. 28, 2020 - Apr. 16, 2021) and Run1 (Nov. 16, 2021 — May 15, 2022), which are divided into subsets based on different electron lifetimes, liquid levels, cathode voltages, and gate voltages, as summarized in Table~\ref{tab:basic_conf}.
The fiducial masses optimized to reject external backgrounds from detector materials are ${2.38 \pm 0.04}$ and $2.48 \pm 0.05$ ton for Run0 and Run1, respectively.

The signal detection efficiency for Run0 (Run1), including reconstruction efficiency, quality cuts efficiency, and ROI efficiency, is estimated by both the data-driven method and waveform-simulation method~\cite{P4SignalModel, P4WaveformSimulation}.
The low threshold of the ROI is 2 photoelectrons (PE) for $Q^c_{S1}$, and the upper limit is 30 keV in $E_{ee}$.
A 99.5\% ER acceptance cut in the space of log$_{10}(Q^c_{S2_b}/Q^c_{S1})$ vs $Q^c_{S1}$ is applied, which can reject more than 66\% (99\%) of the NR events from neutrons (the coherent elastic scattering of solar $^8$B neutrino on xenon nucleus)  and 95\% of NR-like background originating from radon plate-out from the polytetrafluoroethylene reflector surface in the TPC.
The overall efficiency is estimated to be $87.9\pm 4.0\%$ ($86.0 \pm 6.4\%$) for Run0 (Run1), and the energy dependence is shown in Fig.~\ref{fig:eff_qc}. 
Above 7 keV, the efficiency plateaus and the average value is $94.4 \pm 4.2\%$ ($92.6 \pm 6.8\%$) for Run0 (Run1).

Major background sources of the PandaX-4T experiment can be categorized into three types. 
Environmental backgrounds are from cosmic muons, solar neutrinos, and radioisotopes from surrounding rocks in the lab.
External backgrounds originate from detector components such as PMTs. 
Internal backgrounds refer to radioisotopes of xenon and other radio impurities dissolved in the target.

\begin{table*}
    \centering
    \caption{Summary of background expectations and best fit of the background-only hypothesis of each component for Run0 and Run1. Systematic uncertainties from the signal model and detection efficiencies are included in the uncertainties presented.}
    \resizebox{\linewidth}{!}{
    \setlength{\tabcolsep}{7mm}{
    \begin{tabular}{c|cc|cc}
    \toprule[1.5pt]
 Run & \multicolumn{2}{c}{Run0} \vline & \multicolumn{2}{c}{Run1}\\
    \midrule[1pt]
       & Expected & Fitted & Expected & Fitted \\
    \midrule[1pt]
 Tritium  & - & 578.6 $\pm$ 32.5 & - & $118.4 \pm 31.1$ \\
    $^{214}$Pb & 327.2 $\pm$ 18.8 & 328.0 $\pm$ 17.1& $724.0 \pm 61.5$& $700.3 \pm 45.3$\\
    $^{212}$Pb & 57.8 $\pm$ 14.7& $57.3 \pm 14.1$ & $103.3 \pm 26.9$& $96.5 \pm 23.8$ \\
    $^{85}$Kr & 94.2 $\pm$ 47.3 & $87.3 \pm 31.2$& $308.1 \pm 95.2$ & $272.2 \pm 58.9$\\
 Material & 49.4 $\pm$ 3.3 & $49.5 \pm 3.1$ & $111.7 \pm 9.9$ & $105.9 \pm 7.8$\\
    $^{136}$Xe  & 36.9 $\pm$ 2.5 & $36.9 \pm 2.4$ & $66.2 \pm 5.9$ & $62.3 \pm 4.6$ \\
    $^{127}$Xe & 6.1 $\pm$ 0.3 & $6.1 \pm 0.3$ & $0.0 \pm 0.0$ & $0.0 \pm 0.0$\\
    $^{124}$Xe & 2.3 $\pm$ 0.6 & $2.3 \pm 0.6$ & $4.0 \pm 1.1$ & $3.9 \pm 1.1$\\
 Solar $\nu$ & 43.0 $\pm$ 4.6 & $42.9 \pm 4.5$& $76.8 \pm 9.4$& $72.6 \pm 8.1$\\
 Accidental & 7.6 $\pm$ 2.4 & $7.8 \pm 2.1$ & $7.1 \pm 2.3$ & $7.0 \pm 1.9$\\
 Total & - & 1196.6$\pm$32.6& - & $1439.2 \pm 36.2$\\
 \midrule
 Observed & \multicolumn{2}{c}{1197} \vline & \multicolumn{2}{c}{1431} \\

    \bottomrule[1.5pt]
    \end{tabular}
 }
 }
    \label{tab:bkg_fit}
\end{table*}

With 2.4~km rock overburden of the Jinping Mountain, the cosmic muon flux at CJPL is measured to be $\mathcal{O}(10^{-10})\rm{\,cm^{-2}s^{-1}}$~\cite{MuonCJPL}, leading to a negligible muon-induced background in the ROI.
Solar neutrinos produced via proton-proton fusion and $^7$Be electron capture are calculated using standard solar model, three-flavor neutrino oscillation, standard model neutrino-electron scattering, as well as updated treatment of the xenon atomic effects~\cite{NuBKG, CHEN2017656, P4pp_paper}. 
A 10\% uncertainty is used based on Borexino measurements~\cite{BorexinoNeutrinos}.
The concentration of K, Th, and U of the CJPL lab environment are measured at Bq/kg level~\cite{CJPL_BKG_CDEX}.
With at least 4~m of water shielding in each direction~\cite{P4_TDR}, the contribution from surrounding rocks is suppressed to a negligible level.
The concentrations of $^{232}$Th and $^{238}$U in ultra-pure water shielding itself are measured to be $10^{-2}$~ppt (part per trillion) and are expected to contribute $\sim 10^{-4}$events/ton/yr/keV~\cite{JHEPbkg} in the ROI.

The background from $\gamma$ rays of detector materials is calculated based on the ER spectrum in the MeV energy range, where characteristic full-absorption peaks can help constrain the radioactivities of different components \emph{in situ}~\cite{DBD_136}.
The external backgrounds contribute a flat shape in the ROI according to a custom \textsc{geant4}-based simulation package BambooMC~\cite{BambooMC,Geant4}, with the dominant contributions from the stainless steel vessels and PMTs.
The expected background counts in the ROI are $49.4 \pm 3.3$ and $111.7 \pm 9.9$ events for Run0 and Run1, extrapolated from the results of Ref.~\cite{DBD_136} with the help of BambooMC.

Internal backgrounds in the ROI are mainly from $\beta$ decays of the progenies of radon, especially $^{222}$Rn dissolved in LXe. 
The decay chain,
 $\rm ^{222}Rn\rightarrow {^{218}Po}\rightarrow {^{214}Pb}\rightarrow {^{214}Bi}\rightarrow {^{214}Po}\rightarrow {^{210}Pb}$,
practically stops at the long-lived $^{210}$Pb.
The concentration of $^{222}$Rn is $7.1 \pm 0.2$ ($8.7 \pm 0.3 $) $\mu$Bq/kg for Run0 (Run1), determined from the number of $\alpha$ decay events.
The $\beta$ decay rate of $^{214}$Pb cannot be accurately determined from $^{222}$Rn.
A depletion effect, i.e., decreased concentration of longer-lived daughter nuclide compared to the parent in LXe, has been reported~\cite{P2_Rn220_Ma}.
The concentration of $^{214}$Pb for Run0 is $4.5 \pm 0.2 $ $\mu$Bq/kg, derived from the spectral fit in 200-1000 keV~\cite{P4Xe134DBD}.
For Run1, the concentration is estimated to be $5.5 \pm 0.3$ $\mu$Bq/kg, based on the ratio of $^{222}$Rn concentrations relative to Run0.
For $^{220}$Rn decay chain, the dominating background comes from $^{212}$Pb, which is estimated to be $0.28 \pm 0.08$ $\mu$Bq/kg with the depletion effect taken into account, using the level of its daughter $^{212}$Po $\alpha$ decays as the proxy. 

Another $\beta$ emission source is the $^{85}$Kr.
The krypton concentration in Run0 (Run1) is 0.52 $\pm$ 0.27 (0.94 $\pm$ 0.28) ppt, estimated with the coincidence of $^{85}$Kr $\beta$ decay to metastable state $\rm ^{85m}$Rb and the subsequent deexcitation $\gamma$ to ground state $\rm ^{85}$Rb. 
The predominant contribution to the uncertainty is the limited statistics of $\beta$-$\gamma$ coincidence, which are 4 events in Run0 and 12 events in Run1.
Assuming an isotopic abundance of $2 \times 10^{-11}$, 94.2 (308.1) $^{85}$Kr events are in our ROI for Run0 (Run1).

In the Run0 data, tritium events are observed at a concentration of $5 \times 10^{-24}$ mol/mol~\cite{P4CommissonRun}.
The contamination is likely introduced to xenon during the PandaX-II calibration runs~\cite{P2_axion}.
After a dedicated tritium removal campaign after Run0, the tritium contribution in Run1 drops to $0.6 \times 10^{-24}$ mol/mol.
We do not impose any constraint of tritium in our spectral fit. 

Background contributions from xenon isotopes are considered, notably the $^{136}$Xe and $^{124}$Xe, along with cosmogenic $^{127}$Xe.
The isotopic abundance of $^{136}$Xe and $^{124}$Xe are measured to be 8.7\% and 0.1\%.
The half-life of $^{136}$Xe's double beta decay is measured in Ref.~\cite{DBD_136}, leading to an estimated background contribution of $36.9 \pm 2.5$ events for Run0 and $66.2 \pm 5.9$ events for Run1.
Double electron capture events of LL-shell of $^{124}$Xe with an expected energy of approximately 10~keV are estimated based on the theoretical branching ratio and the measured $^{124}$Xe half-life from XENON1T~\cite{XENON:Xe124}.
The expected number of events for Run0 and Run1 is $2.3 \pm 0.6$ and $4.0 \pm 1.1$, respectively. 
The decay from $L$-shell electron capture (5.2 keV) of $^{127}$Xe also contributes to our ROI, which can be determined by the 408 keV peak ($K$-shell capture mixed with the 375 keV $\gamma$-ray from $^{127}$I). 

In addition to the physical background, accidental coincidence (AC) events resulting from the random pairing of isolated $S1$ and $S2$ signals also contribute to the background.  
These spurious events are estimated to be $7.6 \pm 2.4$ and $7.1 \pm 2.3$ in Run0 and Run1, respectively~\cite{Salam_AC}. 
The background levels do not scale with the exposure due to different cut efficiencies in the two runs. 

\begin{figure}[t]
    \centering
    \includegraphics[width=\columnwidth]{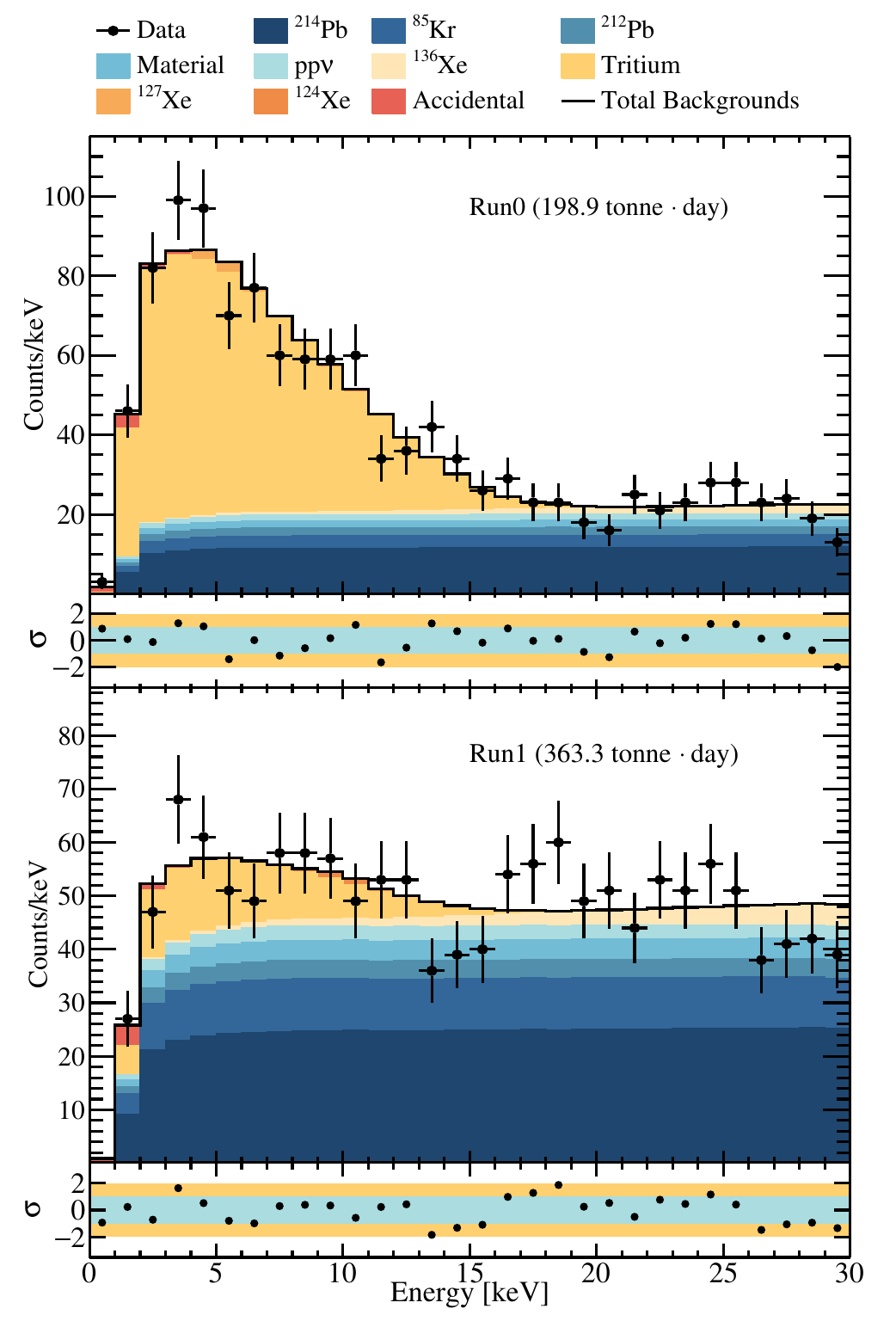}
    \caption{Joint fit of the Run0 (top) and Run1 (bottom) ER data with background-only hypothesis. 
 Backgrounds are stacked in the sequence presented by the legend.
 The 1$\sigma$ and 2$\sigma$ residual bands are depicted with dark cyan and yellow bands, respectively.
 Details of the fit results are listed in Table~\ref{tab:bkg_fit}.}
    \label{fig:bkg_fit}
\end{figure}

Our final analysis immediately followed the unblinding of Run0 and Run1 data in Ref.~\cite{P4_WIMP_Run0Run1}, and no post-unblinding adjustments were made on the event selection cuts and background models.
In total, 1197 (1431) ER events are selected in Run0 (Run1), different from those in Ref.~\cite{P4_WIMP_Run0Run1} only due to a broader range of ROI, and the removal of NR-like data beyond the 99.5\% ER acceptance.
A binned one-dimensional likelihood fit in energy based on HistFitter~\cite{baak_histfitter_2015, Cranmer2012HistFactoryAT} is performed to test the background-only hypotheses.
Background estimates in Table~\ref{tab:bkg_fit} are taken with prior uncertainties as constraints.
Systematics, including those corresponding to the signal model in PandaX-4T~\cite{P4SignalModel} and detection efficiency (see Fig.~\ref{fig:eff_qc}), are also incorporated into the fit as nuisance parameters.
Among all signal model uncertainties, the electron recombination rate (affecting the energy division between scintillation and ionization) impacts the low energy spectrum, as a change in the light yield mostly influences events close to the lower threshold of $Q^c_{S1} > 2$ PE~\cite{LZ_SA}.
The propagated uncertainty on the spectrum is within $\pm$3.7\% below 2~keV and has no impact above.
The best fit of a background-only model is shown in Fig.~\ref{fig:bkg_fit}, with the number of fitted background events summarized in Table~\ref{tab:bkg_fit}.
The fitted background contributions agree with the expected values, with all fitted nuisance parameters within $\pm1\sigma$ of the input values.
The ROI's average fitted ER background rate (including efficiency) is $73.8 \pm 2.0$ and $48.0 \pm 1.2$ events/ton/yr/keV in Run0 and Run1, respectively. 
The data are consistent with the background-only model with a $p$ value of 0.16.

\begin{figure*}[ht]
    \includegraphics[width=2\columnwidth]{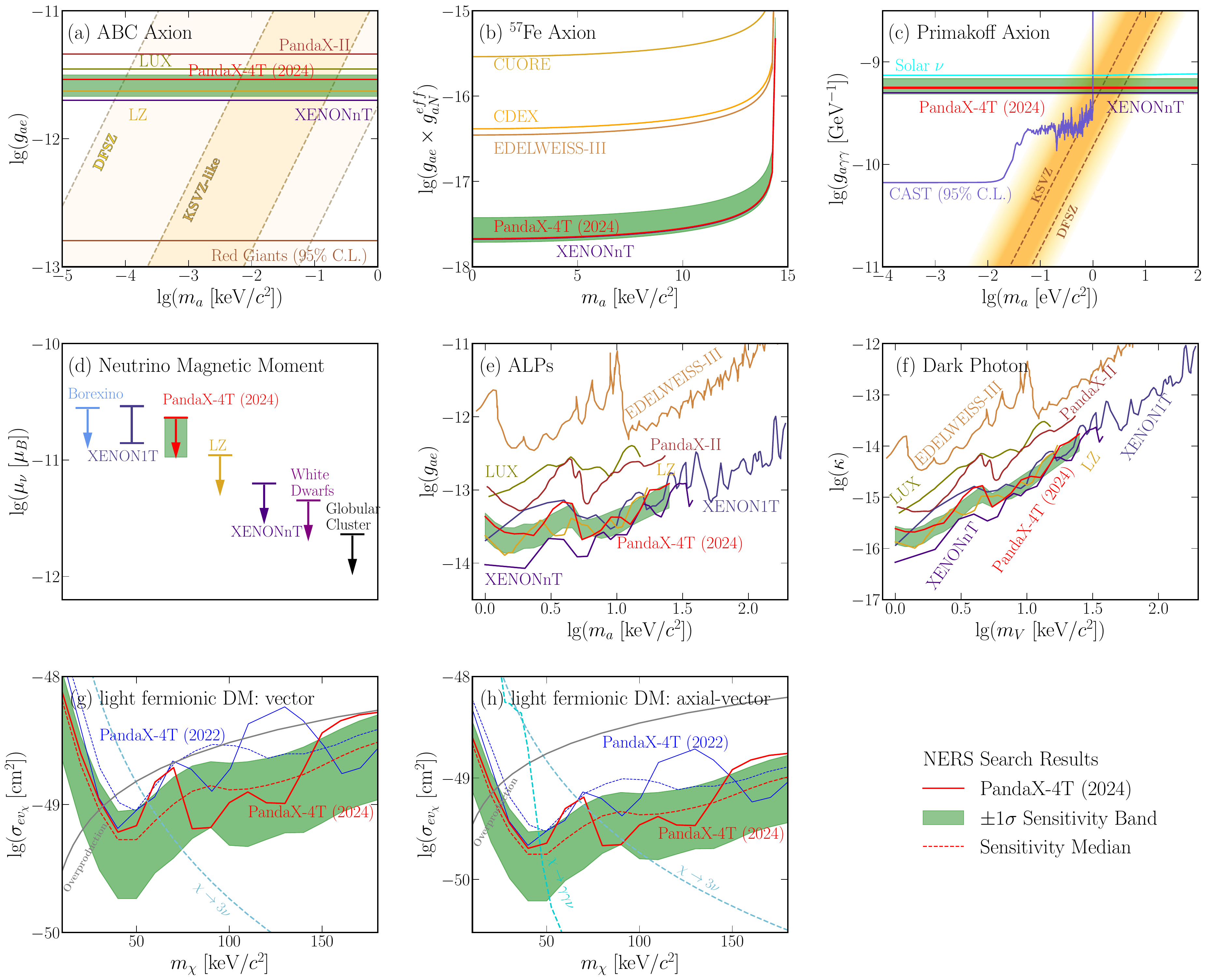}
    \caption{The 90\% CL upper limits of couplings of different NERS. (a) $g_{ae}$ for ABC axion; (b) $g_{ae}g^{\text{eff}}_{aN}$ for $^{57}$Fe axion; (c) $g_{a\gamma\gamma}$ for Primakoff axion; (d) $\mu_{\nu}$ for solar neutrino with anomalous magnetic moment ($\mu_B \equiv e\hbar/2m_e$ is the Bohr magneton); (e) $g_{ae}$ for ALPs; (f) mixing angle $\kappa$ for dark photons; (g)(h) cross sections $\sigma_e v_{\chi}$ of light fermionic DM-$\nu$ conversion for the vector and axial-vector interactions. In (g)(h), the limits (blue solid line) and sensitivity median (blue dashed line) of Run0 published results~\cite{FDM_ZhangDan}, denoted as PandaX-4T (2022), are also overlaid.
 In scenarios (e) to (h), we assume that each type of dark matter candidate constitutes 100\% of the local dark matter.
 The limits from other experiments~\cite{AxionLimits, LUX_axion, RedGiant, Fe57_CUORE, CDEX_axion_2019, EDELWEISS_III_axion, SolarNu_axion, CAST, BorexinoNeutrinos, Xenon1TwhiteDwarfmumag, WhiteDwarfmumag, HXMT, NuStarM31, INTEGRAL_Bouchet_2011, XENON1T_excess, XENONnT_axion, LZ_SA} are also shown.The shaded orange region in (a)(c) represents predictions derived from two benchmark QCD axion models: KSVZ~\cite{KSVZ_1, KSVZ_2} and DFSZ~\cite{DFSZ_1, DFSZ_2}.}
    \label{fig:limit}
\end{figure*}

NERS hypotheses are tested with the background-plus-signal fits using the Run0 and Run1 data.
The best fit of the ABC axions hypothesis yields $11.6^{+24.8}_{-11.6}$ events (ranged to positive only), while the best fits for axions originating from the Primakoff conversion and $^{57}$Fe are null, assuming an axion mass of $1 \times 10^{-5}$ keV/$c^2$.
For the anomalous neutrino magnetic moment hypothesis, the best fit is $103.7^{+114.3}_{-103.7}$ events, also compatible with zero.
ALP signals, as well as those from dark photons and DM-$\nu$ conversion, exhibit a characteristic peak structure that correlates with their mass. 
The mass scanning for ALPs and dark photons is conducted within the range of 1 to 25 keV/$c^2$, while for DM-$\nu$ conversion, the range is from 10 to 180 keV/$c^2$.  
The data exhibit an upward fluctuation in the 3 - 5 keV range for both Run0 and Run1, which results in a local significance of 1.9$\sigma$ for ALPs and dark photons at a mass of 3.3 keV/$c^2$, and 1.6$\sigma$ for light DM at a mass of 60 keV/$c^2$ (peak of recoil energy at 4.1 keV) relative to the background-only hypothesis~\cite{Cowan:2010js}.

As none of the local excess is beyond 3$\sigma$, according to the conventions in Ref.~\cite{Baxter:PLR_conventions}, we proceed with the upper limit interpretation of our data using the profile likelihood ratio approach outlined in Ref.~\cite{Cowan:2010js}.
The upper limits with 90\% confidence level (CL) of axion couplings ($g_{ae},\,g_{ae}g_{aN}^{\text{eff}},\,g_{a\gamma}$), $\mu_\nu$, kinetic mixing of dark-to-SM photon $\kappa$ and DM-$\nu$ conversion cross sections ($\sigma_{\chi e} v_{\chi}$) are shown in Fig.~\ref{fig:limit}.
The corresponding 1$\sigma$ sensitivities are shown as green bands in the figure.
In the scanning over DM mass in the last four panels in Fig.~\ref{fig:limit}, local data fluctuations cause corresponding variations in the derived limits.
For conservativeness, limits due to upward fluctuation of data are reported as they are, but those due to downward fluctuation are power constrained to -1$\sigma$ boundary of the sensitivity band~\cite{cowan2011powerconstrainedlimits}.
For comparison, constraints from other direct detection experiments and astrophysical observations are overlaid.

The upper limit derived from ABC axions is $g_{ae} < 2.9\times10^{-12}$ for $m_a < 1$ keV/$c^2$.
For axions produced by the 14.4 keV $^{57}$Fe transition, $g_{ae}g_{aN}^{\text{eff}}$ are constrained to less than $2.1\times10^{-18}$ for $m_a=1.0\times 10^{-5}$ keV/$c^2$, which is as competitive as the constraints from XENONnT~\cite{XENONnT_axion}.
The axion-photon coupling $g_{a\gamma\gamma}$ is constrained to $5.6\times 10^{-10}$ GeV$^{-1}$ for $m_a < 100$ keV/$c^2$,  which represents one of the most competitive limits on $g_{a\gamma\gamma}$ from a terrestrial experiment, and disfavoring KSVZ~\cite{KSVZ_1, KSVZ_2} and DFSZ~\cite{DFSZ_1, DFSZ_2} axions above a few eV/$c^2$.
The upper limit on the anomalous neutrino magnetic moment is $\mu_{\nu} < 2.3\times10^{-11}\mu_B$.
Compared to our previous results~\cite{P2_axion}, the constraints on $g_{ae}$ and $\mu_{\nu}$ have been improved by about 2 times. 
Our results for ALPs and dark photons limits are slightly weaker than those from XENONnT~\cite{XENONnT_axion}.
Our sensitivities for DM-$\nu$ conversion are notably improved compared to the previous best results in Ref.~\cite{FDM_ZhangDan}, except at $m_\chi$ around 60 keV/$c^2$ and between 150 - 180 keV/$c^2$ due to the upward data fluctuation at around 3 keV and 24 - 25 keV.

In summary, we have searched for NERS, including solar axions, neutrinos with anomalous magnetic moment, ALPs, dark photons, and light fermionic DM-$\nu$ conversion, with the 1.54~$\rm ton \cdot yr$ of PandaX-4T low-energy electronic recoil data. 
No statistically significant NERS have been observed.
Consequently, stringent upper limits are established on NERS. 
In particular, we report leading limits on solar $^{57}$Fe axions and light fermionic DM-$\nu$ conversion, the latter of which is the most stringent result for most of the mass range of 10 - 180 keV/$c^2$.
After a long shutdown due to the construction work in CJPL, PandaX-4T has restarted detector commissioning.
Improved search sensitivities for NERS are expected with new data.

This project is supported in part by grants from National Key R\&D Program of China (No. 2023YFA1606200, No. 2023YFA1606201, National Science Foundation of China (No. 12090060, No. 12090061, No. 12105008, No. U23B2070), and by Office of Science and Technology, Shanghai Municipal Government (Grant No. 21TQ1400218, No. 22JC1410100, No. 23JC1410200, No. ZJ2023-ZD-003). We thank the support by the Fundamental Research Funds for the Central Universities, Hongwen Foundation in Hong Kong, New Cornerstone Science Foundation, Tencent Foundation in China, and Yangyang Development Fund. Finally, we thank the CJPL administration and the Yalong River Hydropower Development Company Ltd. for indispensable logistical support and other help. 
\bibliographystyle{apsrev4-1}
\bibliography{refs.bib}

\begin{thebibliography}{80}%
\makeatletter
\providecommand \@ifxundefined [1]{%
 \@ifx{#1\undefined}
}%
\providecommand \@ifnum [1]{%
 \ifnum #1\expandafter \@firstoftwo
 \else \expandafter \@secondoftwo
 \fi
}%
\providecommand \@ifx [1]{%
 \ifx #1\expandafter \@firstoftwo
 \else \expandafter \@secondoftwo
 \fi
}%
\providecommand \natexlab [1]{#1}%
\providecommand \enquote  [1]{``#1''}%
\providecommand \bibnamefont  [1]{#1}%
\providecommand \bibfnamefont [1]{#1}%
\providecommand \citenamefont [1]{#1}%
\providecommand \href@noop [0]{\@secondoftwo}%
\providecommand \href [0]{\begingroup \@sanitize@url \@href}%
\providecommand \@href[1]{\@@startlink{#1}\@@href}%
\providecommand \@@href[1]{\endgroup#1\@@endlink}%
\providecommand \@sanitize@url [0]{\catcode `\\12\catcode `\$12\catcode `\&12\catcode `\#12\catcode `\^12\catcode `\_12\catcode `\%12\relax}%
\providecommand \@@startlink[1]{}%
\providecommand \@@endlink[0]{}%
\providecommand \url  [0]{\begingroup\@sanitize@url \@url }%
\providecommand \@url [1]{\endgroup\@href {#1}{\urlprefix }}%
\providecommand \urlprefix  [0]{URL }%
\providecommand \Eprint [0]{\href }%
\providecommand \doibase [0]{http://dx.doi.org/}%
\providecommand \selectlanguage [0]{\@gobble}%
\providecommand \bibinfo  [0]{\@secondoftwo}%
\providecommand \bibfield  [0]{\@secondoftwo}%
\providecommand \translation [1]{[#1]}%
\providecommand \BibitemOpen [0]{}%
\providecommand \bibitemStop [0]{}%
\providecommand \bibitemNoStop [0]{.\EOS\space}%
\providecommand \EOS [0]{\spacefactor3000\relax}%
\providecommand \BibitemShut  [1]{\csname bibitem#1\endcsname}%
\let\auto@bib@innerbib\@empty
\bibitem [{\citenamefont {Bo}\ \emph {et~al.}(2024{\natexlab{a}})\citenamefont {Bo} \emph {et~al.}}]{P4_WIMP_Run0Run1}%
  \BibitemOpen
  \bibfield  {author} {\bibinfo {author} {\bibfnamefont {Z.}~\bibnamefont {Bo}} \emph {et~al.} (\bibinfo {collaboration} {PandaX Collaboration}),\ }\href {https://arxiv.org/abs/2408.00664} {\  (\bibinfo {year} {2024}{\natexlab{a}})},\ \Eprint {http://arxiv.org/abs/2408.00664} {arXiv:2408.00664 [hep-ex]} \BibitemShut {NoStop}%
\bibitem [{\citenamefont {Bo}\ \emph {et~al.}(2024{\natexlab{b}})\citenamefont {Bo} \emph {et~al.}}]{P4_B8_Run0Run1}%
  \BibitemOpen
  \bibfield  {author} {\bibinfo {author} {\bibfnamefont {Z.}~\bibnamefont {Bo}} \emph {et~al.} (\bibinfo {collaboration} {PandaX Collaboration}),\ }\href {\doibase 10.1103/PhysRevLett.133.191001} {\bibfield  {journal} {\bibinfo  {journal} {Phys. Rev. Lett.}\ }\textbf {\bibinfo {volume} {133}},\ \bibinfo {pages} {191001} (\bibinfo {year} {2024}{\natexlab{b}})}\BibitemShut {NoStop}%
\bibitem [{\citenamefont {Aprile}\ \emph {et~al.}(2023)\citenamefont {Aprile} \emph {et~al.}}]{XENONnT_WIMP}%
  \BibitemOpen
  \bibfield  {author} {\bibinfo {author} {\bibfnamefont {E.}~\bibnamefont {Aprile}} \emph {et~al.} (\bibinfo {collaboration} {XENON Collaboration}),\ }\href {\doibase 10.1103/PhysRevLett.131.041003} {\bibfield  {journal} {\bibinfo  {journal} {Phys. Rev. Lett.}\ }\textbf {\bibinfo {volume} {131}},\ \bibinfo {pages} {041003} (\bibinfo {year} {2023})}\BibitemShut {NoStop}%
\bibitem [{\citenamefont {Aprile}\ \emph {et~al.}(2024)\citenamefont {Aprile} \emph {et~al.}}]{XENONnT_B8}%
  \BibitemOpen
  \bibfield  {author} {\bibinfo {author} {\bibfnamefont {E.}~\bibnamefont {Aprile}} \emph {et~al.} (\bibinfo {collaboration} {XENON Collaboration}),\ }\href {\doibase 10.1103/PhysRevLett.133.191002} {\bibfield  {journal} {\bibinfo  {journal} {Phys. Rev. Lett.}\ }\textbf {\bibinfo {volume} {133}},\ \bibinfo {pages} {191002} (\bibinfo {year} {2024})}\BibitemShut {NoStop}%
\bibitem [{\citenamefont {Aalbers}\ \emph {et~al.}(2023{\natexlab{a}})\citenamefont {Aalbers} \emph {et~al.}}]{LZ_WIMP}%
  \BibitemOpen
  \bibfield  {author} {\bibinfo {author} {\bibfnamefont {J.}~\bibnamefont {Aalbers}} \emph {et~al.} (\bibinfo {collaboration} {LUX-ZEPLIN Collaboration}),\ }\href {\doibase 10.1103/PhysRevLett.131.041002} {\bibfield  {journal} {\bibinfo  {journal} {Phys. Rev. Lett.}\ }\textbf {\bibinfo {volume} {131}},\ \bibinfo {pages} {041002} (\bibinfo {year} {2023}{\natexlab{a}})}\BibitemShut {NoStop}%
\bibitem [{\citenamefont {Zhou}\ \emph {et~al.}(2021)\citenamefont {Zhou} \emph {et~al.}}]{P2_axion}%
  \BibitemOpen
  \bibfield  {author} {\bibinfo {author} {\bibfnamefont {X.}~\bibnamefont {Zhou}} \emph {et~al.} (\bibinfo {collaboration} {PandaX Collaboration}),\ }\href {\doibase 10.1088/0256-307x/38/1/011301} {\bibfield  {journal} {\bibinfo  {journal} {Chinese Physics Letters}\ }\textbf {\bibinfo {volume} {38}},\ \bibinfo {pages} {011301} (\bibinfo {year} {2021})}\BibitemShut {NoStop}%
\bibitem [{\citenamefont {Zhang}\ \emph {et~al.}(2022)\citenamefont {Zhang} \emph {et~al.}}]{FDM_ZhangDan}%
  \BibitemOpen
  \bibfield  {author} {\bibinfo {author} {\bibfnamefont {D.}~\bibnamefont {Zhang}} \emph {et~al.} (\bibinfo {collaboration} {PandaX Collaboration}),\ }\href {\doibase 10.1103/PhysRevLett.129.161804} {\bibfield  {journal} {\bibinfo  {journal} {Phys. Rev. Lett.}\ }\textbf {\bibinfo {volume} {129}},\ \bibinfo {pages} {161804} (\bibinfo {year} {2022})}\BibitemShut {NoStop}%
\bibitem [{\citenamefont {Aprile}\ \emph {et~al.}(2022{\natexlab{a}})\citenamefont {Aprile} \emph {et~al.}}]{XENONnT_axion}%
  \BibitemOpen
  \bibfield  {author} {\bibinfo {author} {\bibfnamefont {E.}~\bibnamefont {Aprile}} \emph {et~al.} (\bibinfo {collaboration} {XENON Collaboration}),\ }\href {\doibase 10.1103/PhysRevLett.129.161805} {\bibfield  {journal} {\bibinfo  {journal} {Phys. Rev. Lett.}\ }\textbf {\bibinfo {volume} {129}},\ \bibinfo {pages} {161805} (\bibinfo {year} {2022}{\natexlab{a}})}\BibitemShut {NoStop}%
\bibitem [{\citenamefont {Aalbers}\ \emph {et~al.}(2023{\natexlab{b}})\citenamefont {Aalbers} \emph {et~al.}}]{LZ_SA}%
  \BibitemOpen
  \bibfield  {author} {\bibinfo {author} {\bibfnamefont {J.}~\bibnamefont {Aalbers}} \emph {et~al.} (\bibinfo {collaboration} {LZ Collaboration}),\ }\href {\doibase 10.1103/PhysRevD.108.072006} {\bibfield  {journal} {\bibinfo  {journal} {Phys. Rev. D}\ }\textbf {\bibinfo {volume} {108}},\ \bibinfo {pages} {072006} (\bibinfo {year} {2023}{\natexlab{b}})}\BibitemShut {NoStop}%
\bibitem [{\citenamefont {Peccei}\ and\ \citenamefont {Quinn}(1977)}]{PecceiQuinn}%
  \BibitemOpen
  \bibfield  {author} {\bibinfo {author} {\bibfnamefont {R.~D.}\ \bibnamefont {Peccei}}\ and\ \bibinfo {author} {\bibfnamefont {H.~R.}\ \bibnamefont {Quinn}},\ }\href {\doibase 10.1103/PhysRevLett.38.1440} {\bibfield  {journal} {\bibinfo  {journal} {Phys. Rev. Lett.}\ }\textbf {\bibinfo {volume} {38}},\ \bibinfo {pages} {1440} (\bibinfo {year} {1977})}\BibitemShut {NoStop}%
\bibitem [{\citenamefont {Weinberg}(1978)}]{Weinberg}%
  \BibitemOpen
  \bibfield  {author} {\bibinfo {author} {\bibfnamefont {S.}~\bibnamefont {Weinberg}},\ }\href {\doibase 10.1103/PhysRevLett.40.223} {\bibfield  {journal} {\bibinfo  {journal} {Phys. Rev. Lett.}\ }\textbf {\bibinfo {volume} {40}},\ \bibinfo {pages} {223} (\bibinfo {year} {1978})}\BibitemShut {NoStop}%
\bibitem [{\citenamefont {Wilczek}(1978)}]{Wilczek}%
  \BibitemOpen
  \bibfield  {author} {\bibinfo {author} {\bibfnamefont {F.}~\bibnamefont {Wilczek}},\ }\href {\doibase 10.1103/PhysRevLett.40.279} {\bibfield  {journal} {\bibinfo  {journal} {Phys. Rev. Lett.}\ }\textbf {\bibinfo {volume} {40}},\ \bibinfo {pages} {279} (\bibinfo {year} {1978})}\BibitemShut {NoStop}%
\bibitem [{\citenamefont {Redondo}(2013)}]{FluxFromAE}%
  \BibitemOpen
  \bibfield  {author} {\bibinfo {author} {\bibfnamefont {J.}~\bibnamefont {Redondo}},\ }\href {\doibase 10.1088/1475-7516/2013/12/008} {\bibfield  {journal} {\bibinfo  {journal} {J. Cosmol. Astropart. Phys.}\ }\textbf {\bibinfo {volume} {2013}},\ \bibinfo {pages} {008} (\bibinfo {year} {2013})}\BibitemShut {NoStop}%
\bibitem [{\citenamefont {Derevianko}\ \emph {et~al.}(2010)\citenamefont {Derevianko}, \citenamefont {Dzuba}, \citenamefont {Flambaum},\ and\ \citenamefont {Pospelov}}]{AxioElectricEffect}%
  \BibitemOpen
  \bibfield  {author} {\bibinfo {author} {\bibfnamefont {A.}~\bibnamefont {Derevianko}}, \bibinfo {author} {\bibfnamefont {V.~A.}\ \bibnamefont {Dzuba}}, \bibinfo {author} {\bibfnamefont {V.~V.}\ \bibnamefont {Flambaum}}, \ and\ \bibinfo {author} {\bibfnamefont {M.}~\bibnamefont {Pospelov}},\ }\href {\doibase 10.1103/PhysRevD.82.065006} {\bibfield  {journal} {\bibinfo  {journal} {Phys. Rev. D}\ }\textbf {\bibinfo {volume} {82}},\ \bibinfo {pages} {065006} (\bibinfo {year} {2010})}\BibitemShut {NoStop}%
\bibitem [{\citenamefont {Primakoff}(1951)}]{Primakoff}%
  \BibitemOpen
  \bibfield  {author} {\bibinfo {author} {\bibfnamefont {H.}~\bibnamefont {Primakoff}},\ }\href {\doibase 10.1103/PhysRev.81.899} {\bibfield  {journal} {\bibinfo  {journal} {Phys. Rev.}\ }\textbf {\bibinfo {volume} {81}},\ \bibinfo {pages} {899} (\bibinfo {year} {1951})}\BibitemShut {NoStop}%
\bibitem [{\citenamefont {Moriyama}(1995)}]{Fe57Proposal}%
  \BibitemOpen
  \bibfield  {author} {\bibinfo {author} {\bibfnamefont {S.}~\bibnamefont {Moriyama}},\ }\href {\doibase 10.1103/PhysRevLett.75.3222} {\bibfield  {journal} {\bibinfo  {journal} {Phys. Rev. Lett.}\ }\textbf {\bibinfo {volume} {75}},\ \bibinfo {pages} {3222} (\bibinfo {year} {1995})}\BibitemShut {NoStop}%
\bibitem [{\citenamefont {Armengaud}\ \emph {et~al.}(2018)\citenamefont {Armengaud} \emph {et~al.}}]{EDELWEISS_III_axion}%
  \BibitemOpen
  \bibfield  {author} {\bibinfo {author} {\bibfnamefont {E.}~\bibnamefont {Armengaud}} \emph {et~al.} (\bibinfo {collaboration} {EDELWEISS Collaboration}),\ }\href {\doibase 10.1103/PhysRevD.98.082004} {\bibfield  {journal} {\bibinfo  {journal} {Phys. Rev. D}\ }\textbf {\bibinfo {volume} {98}},\ \bibinfo {pages} {082004} (\bibinfo {year} {2018})}\BibitemShut {NoStop}%
\bibitem [{\citenamefont {Giunti}\ and\ \citenamefont {Studenikin}(2015)}]{Giunti:2014ixa}%
  \BibitemOpen
  \bibfield  {author} {\bibinfo {author} {\bibfnamefont {C.}~\bibnamefont {Giunti}}\ and\ \bibinfo {author} {\bibfnamefont {A.}~\bibnamefont {Studenikin}},\ }\href {\doibase 10.1103/RevModPhys.87.531} {\bibfield  {journal} {\bibinfo  {journal} {Rev. Mod. Phys.}\ }\textbf {\bibinfo {volume} {87}},\ \bibinfo {pages} {531} (\bibinfo {year} {2015})},\ \Eprint {http://arxiv.org/abs/1403.6344} {arXiv:1403.6344 [hep-ph]} \BibitemShut {NoStop}%
\bibitem [{\citenamefont {Holdom}(1986)}]{DarkPhoton}%
  \BibitemOpen
  \bibfield  {author} {\bibinfo {author} {\bibfnamefont {B.}~\bibnamefont {Holdom}},\ }\href {\doibase https://doi.org/10.1016/0370-2693(86)91377-8} {\bibfield  {journal} {\bibinfo  {journal} {Phys. Lett. B}\ }\textbf {\bibinfo {volume} {166}},\ \bibinfo {pages} {196} (\bibinfo {year} {1986})}\BibitemShut {NoStop}%
\bibitem [{\citenamefont {Dror}\ \emph {et~al.}(2020)\citenamefont {Dror}, \citenamefont {Elor},\ and\ \citenamefont {McGehee}}]{Dror_2020}%
  \BibitemOpen
  \bibfield  {author} {\bibinfo {author} {\bibfnamefont {J.~A.}\ \bibnamefont {Dror}}, \bibinfo {author} {\bibfnamefont {G.}~\bibnamefont {Elor}}, \ and\ \bibinfo {author} {\bibfnamefont {R.}~\bibnamefont {McGehee}},\ }\href {\doibase 10.1103/physrevlett.124.181301} {\bibfield  {journal} {\bibinfo  {journal} {Phys. Rev. Lett.}\ }\textbf {\bibinfo {volume} {124}} (\bibinfo {year} {2020}),\ 10.1103/physrevlett.124.181301}\BibitemShut {NoStop}%
\bibitem [{\citenamefont {Dror}\ \emph {et~al.}(2021)\citenamefont {Dror}, \citenamefont {Elor}, \citenamefont {McGehee},\ and\ \citenamefont {Yu}}]{FDM_Dror}%
  \BibitemOpen
  \bibfield  {author} {\bibinfo {author} {\bibfnamefont {J.~A.}\ \bibnamefont {Dror}}, \bibinfo {author} {\bibfnamefont {G.}~\bibnamefont {Elor}}, \bibinfo {author} {\bibfnamefont {R.}~\bibnamefont {McGehee}}, \ and\ \bibinfo {author} {\bibfnamefont {T.-T.}\ \bibnamefont {Yu}},\ }\href {\doibase 10.1103/PhysRevD.103.035001} {\bibfield  {journal} {\bibinfo  {journal} {Phys. Rev. D}\ }\textbf {\bibinfo {volume} {103}},\ \bibinfo {pages} {035001} (\bibinfo {year} {2021})}\BibitemShut {NoStop}%
\bibitem [{\citenamefont {Ge}\ \emph {et~al.}(2022{\natexlab{a}})\citenamefont {Ge}, \citenamefont {He}, \citenamefont {Ma},\ and\ \citenamefont {Sheng}}]{geRevisitingFermionicDark2022}%
  \BibitemOpen
  \bibfield  {author} {\bibinfo {author} {\bibfnamefont {S.-F.}\ \bibnamefont {Ge}}, \bibinfo {author} {\bibfnamefont {X.-G.}\ \bibnamefont {He}}, \bibinfo {author} {\bibfnamefont {X.-D.}\ \bibnamefont {Ma}}, \ and\ \bibinfo {author} {\bibfnamefont {J.}~\bibnamefont {Sheng}},\ }\href {https://link.springer.com/article/10.1007/JHEP05%282022%29191} {\bibfield  {journal} {\bibinfo  {journal} {J. High Energy Phys.}\ }\textbf {\bibinfo {volume} {2022}},\ \bibinfo {pages} {191} (\bibinfo {year} {2022}{\natexlab{a}})}\BibitemShut {NoStop}%
\bibitem [{sup()}]{supp}%
  \BibitemOpen
  \href@noop {} {}\bibinfo {note} {See Supplemental Material at \href{https://journals.aps.org/prl/supplemental/10.1103/PhysRevLett.134.041001/Supplement_Exploring_New_Physics_with_PandaX_4T_Low_Energy_Electronic_R.pdf}{https://journals.aps.org/ prl/supplemental/10.1103/PhysRevLett.134.041001/ Supplement\_Exploring\_New\_Physics\_with\_PandaX\_ 4T\_Low\_Energy\_Electronic\_R.pdf} for details of the production and detection of solar axions, neutrinos with anomalous magnetic moment, axionlike particles, dark photons, and light fermionic DM-$\nu$ conversion, which includes Refs. [24-40].}\BibitemShut {Stop}%
\bibitem [{\citenamefont {Barth}\ \emph {et~al.}(2013)\citenamefont {Barth} \emph {et~al.}}]{CAST_2013}%
  \BibitemOpen
  \bibfield  {author} {\bibinfo {author} {\bibfnamefont {K.}~\bibnamefont {Barth}} \emph {et~al.},\ }\href {\doibase 10.1088/1475-7516/2013/05/010} {\bibfield  {journal} {\bibinfo  {journal} {J. Cosmol. Astropart. Phys.}\ }\textbf {\bibinfo {volume} {2013}},\ \bibinfo {pages} {010} (\bibinfo {year} {2013})}\BibitemShut {NoStop}%
\bibitem [{\citenamefont {Veigele}(1973)}]{pe_xsec}%
  \BibitemOpen
  \bibfield  {author} {\bibinfo {author} {\bibfnamefont {W.}~\bibnamefont {Veigele}},\ }\href {\doibase https://doi.org/10.1016/S0092-640X(73)80015-4} {\bibfield  {journal} {\bibinfo  {journal} {Atomic Data and Nuclear Data Tables}\ }\textbf {\bibinfo {volume} {5}},\ \bibinfo {pages} {51} (\bibinfo {year} {1973})}\BibitemShut {NoStop}%
\bibitem [{\citenamefont {Wu}\ \emph {et~al.}(2023)\citenamefont {Wu}, \citenamefont {Liu}, \citenamefont {C.}, \citenamefont {Singh}, \citenamefont {Chen}, \citenamefont {Chi}, \citenamefont {Pandey},\ and\ \citenamefont {Wong}}]{IP_3channels}%
  \BibitemOpen
  \bibfield  {author} {\bibinfo {author} {\bibfnamefont {C.-P.}\ \bibnamefont {Wu}}, \bibinfo {author} {\bibfnamefont {C.-P.}\ \bibnamefont {Liu}}, \bibinfo {author} {\bibfnamefont {G.}~\bibnamefont {C.}}, \bibinfo {author} {\bibfnamefont {L.}~\bibnamefont {Singh}}, \bibinfo {author} {\bibfnamefont {J.-W.}\ \bibnamefont {Chen}}, \bibinfo {author} {\bibfnamefont {H.-C.}\ \bibnamefont {Chi}}, \bibinfo {author} {\bibfnamefont {M.~K.}\ \bibnamefont {Pandey}}, \ and\ \bibinfo {author} {\bibfnamefont {H.~T.}\ \bibnamefont {Wong}},\ }\href {\doibase 10.1103/PhysRevD.108.043029} {\bibfield  {journal} {\bibinfo  {journal} {Phys. Rev. D}\ }\textbf {\bibinfo {volume} {108}},\ \bibinfo {pages} {043029} (\bibinfo {year} {2023})}\BibitemShut {NoStop}%
\bibitem [{\citenamefont {Buchmüller}\ and\ \citenamefont {Hoogeveen}(1990)}]{Buch_BraggScat}%
  \BibitemOpen
  \bibfield  {author} {\bibinfo {author} {\bibfnamefont {W.}~\bibnamefont {Buchmüller}}\ and\ \bibinfo {author} {\bibfnamefont {F.}~\bibnamefont {Hoogeveen}},\ }\href {\doibase https://doi.org/10.1016/0370-2693(90)91444-G} {\bibfield  {journal} {\bibinfo  {journal} {Phys. Lett. B}\ }\textbf {\bibinfo {volume} {237}},\ \bibinfo {pages} {278} (\bibinfo {year} {1990})}\BibitemShut {NoStop}%
\bibitem [{\citenamefont {Creswick}\ \emph {et~al.}(1998)\citenamefont {Creswick}, \citenamefont {{Avignone III}}, \citenamefont {Farach}, \citenamefont {Collar}, \citenamefont {Gattone}, \citenamefont {Nussinov},\ and\ \citenamefont {Zioutas}}]{IPE_Theory_Ge}%
  \BibitemOpen
  \bibfield  {author} {\bibinfo {author} {\bibfnamefont {R.}~\bibnamefont {Creswick}}, \bibinfo {author} {\bibfnamefont {F.}~\bibnamefont {{Avignone III}}}, \bibinfo {author} {\bibfnamefont {H.}~\bibnamefont {Farach}}, \bibinfo {author} {\bibfnamefont {J.}~\bibnamefont {Collar}}, \bibinfo {author} {\bibfnamefont {A.}~\bibnamefont {Gattone}}, \bibinfo {author} {\bibfnamefont {S.}~\bibnamefont {Nussinov}}, \ and\ \bibinfo {author} {\bibfnamefont {K.}~\bibnamefont {Zioutas}},\ }\href {\doibase https://doi.org/10.1016/S0370-2693(98)00183-X} {\bibfield  {journal} {\bibinfo  {journal} {Phys. Lett. B}\ }\textbf {\bibinfo {volume} {427}},\ \bibinfo {pages} {235} (\bibinfo {year} {1998})}\BibitemShut {NoStop}%
\bibitem [{\citenamefont {Brown}\ \emph {et~al.}(2006)\citenamefont {Brown}, \citenamefont {Fox}, \citenamefont {Maslen}, \citenamefont {O'Keefe},\ and\ \citenamefont {Willis}}]{RHFFormFactor}%
  \BibitemOpen
  \bibfield  {author} {\bibinfo {author} {\bibfnamefont {P.}~\bibnamefont {Brown}}, \bibinfo {author} {\bibfnamefont {A.}~\bibnamefont {Fox}}, \bibinfo {author} {\bibfnamefont {E.}~\bibnamefont {Maslen}}, \bibinfo {author} {\bibfnamefont {M.}~\bibnamefont {O'Keefe}}, \ and\ \bibinfo {author} {\bibfnamefont {B.}~\bibnamefont {Willis}},\ }\href {\doibase https://doi.org/10.1107/97809553602060000600} {\bibfield  {journal} {\bibinfo  {journal} {American Cancer Society}\ }\textbf {\bibinfo {volume} {6.1}},\ \bibinfo {pages} {554} (\bibinfo {year} {2006})}\BibitemShut {NoStop}%
\bibitem [{\citenamefont {Doyle}\ and\ \citenamefont {Turner}(1968)}]{RHF_Xray}%
  \BibitemOpen
  \bibfield  {author} {\bibinfo {author} {\bibfnamefont {P.}~\bibnamefont {Doyle}}\ and\ \bibinfo {author} {\bibfnamefont {P.}~\bibnamefont {Turner}},\ }\href {\doibase 10.1107/S0567739468000756} {\bibfield  {journal} {\bibinfo  {journal} {Acta Crystallographica Section A}\ }\textbf {\bibinfo {volume} {24}},\ \bibinfo {pages} {390 – 397} (\bibinfo {year} {1968})},\ \bibinfo {note} {cited by: 1936}\BibitemShut {NoStop}%
\bibitem [{\citenamefont {Abe}\ \emph {et~al.}(2021)\citenamefont {Abe}, \citenamefont {Hamaguchi},\ and\ \citenamefont {Nagata}}]{IPE_RHFFormFactor}%
  \BibitemOpen
  \bibfield  {author} {\bibinfo {author} {\bibfnamefont {T.}~\bibnamefont {Abe}}, \bibinfo {author} {\bibfnamefont {K.}~\bibnamefont {Hamaguchi}}, \ and\ \bibinfo {author} {\bibfnamefont {N.}~\bibnamefont {Nagata}},\ }\href {\doibase https://doi.org/10.1016/j.physletb.2021.136174} {\bibfield  {journal} {\bibinfo  {journal} {Phys. Lett. B}\ }\textbf {\bibinfo {volume} {815}},\ \bibinfo {pages} {136174} (\bibinfo {year} {2021})}\BibitemShut {NoStop}%
\bibitem [{\citenamefont {Wong}(2005)}]{Wong:2004sp}%
  \BibitemOpen
  \bibfield  {author} {\bibinfo {author} {\bibfnamefont {H.~T.}\ \bibnamefont {Wong}},\ }\href {\doibase 10.1016/j.nuclphysbps.2005.01.106} {\bibfield  {journal} {\bibinfo  {journal} {Nucl. Phys. B Proc. Suppl.}\ }\textbf {\bibinfo {volume} {143}},\ \bibinfo {pages} {205} (\bibinfo {year} {2005})},\ \Eprint {http://arxiv.org/abs/hep-ex/0409003} {arXiv:hep-ex/0409003} \BibitemShut {NoStop}%
\bibitem [{\citenamefont {Beacom}\ and\ \citenamefont {Vogel}(1999)}]{Beacom:1999prl}%
  \BibitemOpen
  \bibfield  {author} {\bibinfo {author} {\bibfnamefont {J.~F.}\ \bibnamefont {Beacom}}\ and\ \bibinfo {author} {\bibfnamefont {P.}~\bibnamefont {Vogel}},\ }\href {\doibase 10.1103/PhysRevLett.83.5222} {\bibfield  {journal} {\bibinfo  {journal} {Phys. Rev. Lett.}\ }\textbf {\bibinfo {volume} {83}},\ \bibinfo {pages} {5222} (\bibinfo {year} {1999})}\BibitemShut {NoStop}%
\bibitem [{\citenamefont {Voloshin}(2010)}]{NuMagAtomicEff1}%
  \BibitemOpen
  \bibfield  {author} {\bibinfo {author} {\bibfnamefont {M.~B.}\ \bibnamefont {Voloshin}},\ }\href {\doibase 10.1103/PhysRevLett.105.201801} {\bibfield  {journal} {\bibinfo  {journal} {Phys. Rev. Lett.}\ }\textbf {\bibinfo {volume} {105}},\ \bibinfo {pages} {201801} (\bibinfo {year} {2010})}\BibitemShut {NoStop}%
\bibitem [{\citenamefont {Kouzakov}\ \emph {et~al.}(2011)\citenamefont {Kouzakov}, \citenamefont {Studenikin},\ and\ \citenamefont {Voloshin}}]{NuMagAtomicEff2}%
  \BibitemOpen
  \bibfield  {author} {\bibinfo {author} {\bibfnamefont {K.~A.}\ \bibnamefont {Kouzakov}}, \bibinfo {author} {\bibfnamefont {A.~I.}\ \bibnamefont {Studenikin}}, \ and\ \bibinfo {author} {\bibfnamefont {M.~B.}\ \bibnamefont {Voloshin}},\ }\href {\doibase 10.1103/PhysRevD.83.113001} {\bibfield  {journal} {\bibinfo  {journal} {Phys. Rev. D}\ }\textbf {\bibinfo {volume} {83}},\ \bibinfo {pages} {113001} (\bibinfo {year} {2011})}\BibitemShut {NoStop}%
\bibitem [{\citenamefont {Hsieh}\ \emph {et~al.}(2019)\citenamefont {Hsieh}, \citenamefont {Singh}, \citenamefont {Wu}, \citenamefont {Chen}, \citenamefont {Chi}, \citenamefont {Liu}, \citenamefont {Pandey},\ and\ \citenamefont {Wong}}]{NuMagAtomicEff3}%
  \BibitemOpen
  \bibfield  {author} {\bibinfo {author} {\bibfnamefont {C.-C.}\ \bibnamefont {Hsieh}}, \bibinfo {author} {\bibfnamefont {L.}~\bibnamefont {Singh}}, \bibinfo {author} {\bibfnamefont {C.-P.}\ \bibnamefont {Wu}}, \bibinfo {author} {\bibfnamefont {J.-W.}\ \bibnamefont {Chen}}, \bibinfo {author} {\bibfnamefont {H.-C.}\ \bibnamefont {Chi}}, \bibinfo {author} {\bibfnamefont {C.-P.}\ \bibnamefont {Liu}}, \bibinfo {author} {\bibfnamefont {M.~K.}\ \bibnamefont {Pandey}}, \ and\ \bibinfo {author} {\bibfnamefont {H.~T.}\ \bibnamefont {Wong}},\ }\href {\doibase 10.1103/PhysRevD.100.073001} {\bibfield  {journal} {\bibinfo  {journal} {Phys. Rev. D}\ }\textbf {\bibinfo {volume} {100}},\ \bibinfo {pages} {073001} (\bibinfo {year} {2019})}\BibitemShut {NoStop}%
\bibitem [{\citenamefont {Pospelov}\ \emph {et~al.}(2008)\citenamefont {Pospelov}, \citenamefont {Ritz},\ and\ \citenamefont {Voloshin}}]{ALP_Pospelov}%
  \BibitemOpen
  \bibfield  {author} {\bibinfo {author} {\bibfnamefont {M.}~\bibnamefont {Pospelov}}, \bibinfo {author} {\bibfnamefont {A.}~\bibnamefont {Ritz}}, \ and\ \bibinfo {author} {\bibfnamefont {M.}~\bibnamefont {Voloshin}},\ }\href {\doibase 10.1103/PhysRevD.78.115012} {\bibfield  {journal} {\bibinfo  {journal} {Phys. Rev. D}\ }\textbf {\bibinfo {volume} {78}},\ \bibinfo {pages} {115012} (\bibinfo {year} {2008})}\BibitemShut {NoStop}%
\bibitem [{\citenamefont {An}\ \emph {et~al.}(2015)\citenamefont {An}, \citenamefont {Pospelov}, \citenamefont {Pradler},\ and\ \citenamefont {Ritz}}]{AN2015331}%
  \BibitemOpen
  \bibfield  {author} {\bibinfo {author} {\bibfnamefont {H.}~\bibnamefont {An}}, \bibinfo {author} {\bibfnamefont {M.}~\bibnamefont {Pospelov}}, \bibinfo {author} {\bibfnamefont {J.}~\bibnamefont {Pradler}}, \ and\ \bibinfo {author} {\bibfnamefont {A.}~\bibnamefont {Ritz}},\ }\href {\doibase https://doi.org/10.1016/j.physletb.2015.06.018} {\bibfield  {journal} {\bibinfo  {journal} {Physics Letters B}\ }\textbf {\bibinfo {volume} {747}},\ \bibinfo {pages} {331} (\bibinfo {year} {2015})}\BibitemShut {NoStop}%
\bibitem [{\citenamefont {Ge}\ \emph {et~al.}(2022{\natexlab{b}})\citenamefont {Ge}, \citenamefont {Pasquini},\ and\ \citenamefont {Sheng}}]{geActiveSterileConversion}%
  \BibitemOpen
  \bibfield  {author} {\bibinfo {author} {\bibfnamefont {S.-F.}\ \bibnamefont {Ge}}, \bibinfo {author} {\bibfnamefont {P.}~\bibnamefont {Pasquini}}, \ and\ \bibinfo {author} {\bibfnamefont {J.}~\bibnamefont {Sheng}},\ }\href {\doibase https://doi.org/10.1007/JHEP05(2022)088} {\bibfield  {journal} {\bibinfo  {journal} {J. High Energy Phys.}\ }\textbf {\bibinfo {volume} {2022}} (\bibinfo {year} {2022}{\natexlab{b}}),\ https://doi.org/10.1007/JHEP05(2022)088}\BibitemShut {NoStop}%
\bibitem [{\citenamefont {Catena}\ \emph {et~al.}(2020)\citenamefont {Catena}, \citenamefont {Emken}, \citenamefont {Spaldin},\ and\ \citenamefont {Tarantino}}]{Kfactor}%
  \BibitemOpen
  \bibfield  {author} {\bibinfo {author} {\bibfnamefont {R.}~\bibnamefont {Catena}}, \bibinfo {author} {\bibfnamefont {T.}~\bibnamefont {Emken}}, \bibinfo {author} {\bibfnamefont {N.~A.}\ \bibnamefont {Spaldin}}, \ and\ \bibinfo {author} {\bibfnamefont {W.}~\bibnamefont {Tarantino}},\ }\href {\doibase 10.1103/PhysRevResearch.2.033195} {\bibfield  {journal} {\bibinfo  {journal} {Phys. Rev. Res.}\ }\textbf {\bibinfo {volume} {2}},\ \bibinfo {pages} {033195} (\bibinfo {year} {2020})}\BibitemShut {NoStop}%
\bibitem [{\citenamefont {Luo}\ \emph {et~al.}(2024)\citenamefont {Luo} \emph {et~al.}}]{P4SignalModel}%
  \BibitemOpen
  \bibfield  {author} {\bibinfo {author} {\bibfnamefont {Y.}~\bibnamefont {Luo}} \emph {et~al.} (\bibinfo {collaboration} {PandaX Collaboration}),\ }\href {\doibase 10.1103/PhysRevD.110.023029} {\bibfield  {journal} {\bibinfo  {journal} {Phys. Rev. D}\ }\textbf {\bibinfo {volume} {110}},\ \bibinfo {pages} {023029} (\bibinfo {year} {2024})}\BibitemShut {NoStop}%
\bibitem [{\citenamefont {Zhang}\ \emph {et~al.}(2019)\citenamefont {Zhang} \emph {et~al.}}]{P4_TDR}%
  \BibitemOpen
  \bibfield  {author} {\bibinfo {author} {\bibfnamefont {H.}~\bibnamefont {Zhang}} \emph {et~al.} (\bibinfo {collaboration} {PandaX Collaboration}),\ }\href {\doibase 10.1007/s11433-018-9259-0} {\bibfield  {journal} {\bibinfo  {journal} {Sci. China Phys. Mech. Astron.}\ }\textbf {\bibinfo {volume} {62}},\ \bibinfo {pages} {31011} (\bibinfo {year} {2019})}\BibitemShut {NoStop}%
\bibitem [{\citenamefont {Li}\ \emph {et~al.}(2024)\citenamefont {Li} \emph {et~al.}}]{P4WaveformSimulation}%
  \BibitemOpen
  \bibfield  {author} {\bibinfo {author} {\bibfnamefont {J.}~\bibnamefont {Li}} \emph {et~al.} (\bibinfo {collaboration} {PandaX Collaboration}),\ }\href {\doibase 10.1088/1674-1137/ad380f} {\bibfield  {journal} {\bibinfo  {journal} {Chinese Phys. C}\ }\textbf {\bibinfo {volume} {48}},\ \bibinfo {pages} {073001} (\bibinfo {year} {2024})}\BibitemShut {NoStop}%
\bibitem [{\citenamefont {Guo}\ \emph {et~al.}(2021)\citenamefont {Guo} \emph {et~al.}}]{MuonCJPL}%
  \BibitemOpen
  \bibfield  {author} {\bibinfo {author} {\bibfnamefont {Z.}~\bibnamefont {Guo}} \emph {et~al.},\ }\href {\doibase 10.1088/1674-1137/abccae} {\bibfield  {journal} {\bibinfo  {journal} {Chinese Phys. C}\ }\textbf {\bibinfo {volume} {45}},\ \bibinfo {pages} {025001} (\bibinfo {year} {2021})}\BibitemShut {NoStop}%
\bibitem [{\citenamefont {Billard}\ \emph {et~al.}(2014)\citenamefont {Billard}, \citenamefont {Figueroa-Feliciano},\ and\ \citenamefont {Strigari}}]{NuBKG}%
  \BibitemOpen
  \bibfield  {author} {\bibinfo {author} {\bibfnamefont {J.}~\bibnamefont {Billard}}, \bibinfo {author} {\bibfnamefont {E.}~\bibnamefont {Figueroa-Feliciano}}, \ and\ \bibinfo {author} {\bibfnamefont {L.}~\bibnamefont {Strigari}},\ }\href {\doibase 10.1103/PhysRevD.89.023524} {\bibfield  {journal} {\bibinfo  {journal} {Phys. Rev. D}\ }\textbf {\bibinfo {volume} {89}},\ \bibinfo {pages} {023524} (\bibinfo {year} {2014})}\BibitemShut {NoStop}%
\bibitem [{\citenamefont {Chen}\ \emph {et~al.}(2017)\citenamefont {Chen}, \citenamefont {Chi}, \citenamefont {Liu},\ and\ \citenamefont {Wu}}]{CHEN2017656}%
  \BibitemOpen
  \bibfield  {author} {\bibinfo {author} {\bibfnamefont {J.-W.}\ \bibnamefont {Chen}}, \bibinfo {author} {\bibfnamefont {H.-C.}\ \bibnamefont {Chi}}, \bibinfo {author} {\bibfnamefont {C.-P.}\ \bibnamefont {Liu}}, \ and\ \bibinfo {author} {\bibfnamefont {C.-P.}\ \bibnamefont {Wu}},\ }\href {\doibase https://doi.org/10.1016/j.physletb.2017.10.029} {\bibfield  {journal} {\bibinfo  {journal} {Phys. Lett. B}\ }\textbf {\bibinfo {volume} {774}},\ \bibinfo {pages} {656} (\bibinfo {year} {2017})}\BibitemShut {NoStop}%
\bibitem [{\citenamefont {Lu}\ \emph {et~al.}(2024)\citenamefont {Lu} \emph {et~al.}}]{P4pp_paper}%
  \BibitemOpen
  \bibfield  {author} {\bibinfo {author} {\bibfnamefont {X.}~\bibnamefont {Lu}} \emph {et~al.} (\bibinfo {collaboration} {PandaX Collaboration}),\ }\href {\doibase 10.1088/1674-1137/ad582a} {\bibfield  {journal} {\bibinfo  {journal} {Chinese Phys. C}\ }\textbf {\bibinfo {volume} {48}},\ \bibinfo {pages} {091001} (\bibinfo {year} {2024})}\BibitemShut {NoStop}%
\bibitem [{\citenamefont {Kumaran}\ \emph {et~al.}(2021)\citenamefont {Kumaran}, \citenamefont {Ludhova}, \citenamefont {Penek},\ and\ \citenamefont {Settanta}}]{BorexinoNeutrinos}%
  \BibitemOpen
  \bibfield  {author} {\bibinfo {author} {\bibfnamefont {S.}~\bibnamefont {Kumaran}}, \bibinfo {author} {\bibfnamefont {L.}~\bibnamefont {Ludhova}}, \bibinfo {author} {\bibfnamefont {O.}~\bibnamefont {Penek}}, \ and\ \bibinfo {author} {\bibfnamefont {G.}~\bibnamefont {Settanta}},\ }\href {https://www.mdpi.com/2218-1997/7/7/231} {\bibfield  {journal} {\bibinfo  {journal} {Universe}\ }\textbf {\bibinfo {volume} {7}} (\bibinfo {year} {2021})}\BibitemShut {NoStop}%
\bibitem [{\citenamefont {Ma}\ \emph {et~al.}(2021)\citenamefont {Ma}, \citenamefont {She}, \citenamefont {Zeng}, \citenamefont {Zeng}, \citenamefont {Jing}, \citenamefont {Yue}, \citenamefont {Cheng}, \citenamefont {Li},\ and\ \citenamefont {Zhang}}]{CJPL_BKG_CDEX}%
  \BibitemOpen
  \bibfield  {author} {\bibinfo {author} {\bibfnamefont {H.}~\bibnamefont {Ma}}, \bibinfo {author} {\bibfnamefont {Z.}~\bibnamefont {She}}, \bibinfo {author} {\bibfnamefont {W.}~\bibnamefont {Zeng}}, \bibinfo {author} {\bibfnamefont {Z.}~\bibnamefont {Zeng}}, \bibinfo {author} {\bibfnamefont {M.}~\bibnamefont {Jing}}, \bibinfo {author} {\bibfnamefont {Q.}~\bibnamefont {Yue}}, \bibinfo {author} {\bibfnamefont {J.}~\bibnamefont {Cheng}}, \bibinfo {author} {\bibfnamefont {J.}~\bibnamefont {Li}}, \ and\ \bibinfo {author} {\bibfnamefont {H.}~\bibnamefont {Zhang}},\ }\href {\doibase https://doi.org/10.1016/j.astropartphys.2021.102560} {\bibfield  {journal} {\bibinfo  {journal} {Astropart. Phys.}\ }\textbf {\bibinfo {volume} {128}},\ \bibinfo {pages} {102560} (\bibinfo {year} {2021})}\BibitemShut {NoStop}%
\bibitem [{\citenamefont {Qian}\ \emph {et~al.}(2022)\citenamefont {Qian} \emph {et~al.}}]{JHEPbkg}%
  \BibitemOpen
  \bibfield  {author} {\bibinfo {author} {\bibfnamefont {Z.}~\bibnamefont {Qian}} \emph {et~al.} (\bibinfo {collaboration} {PandaX Collaboration}),\ }\href {\doibase 10.1007/JHEP06(2022)147} {\bibfield  {journal} {\bibinfo  {journal} {J. High Energy Phys.}\ }\textbf {\bibinfo {volume} {2022}},\ \bibinfo {pages} {147} (\bibinfo {year} {2022})}\BibitemShut {NoStop}%
\bibitem [{\citenamefont {Si}\ \emph {et~al.}(2022)\citenamefont {Si} \emph {et~al.}}]{DBD_136}%
  \BibitemOpen
  \bibfield  {author} {\bibinfo {author} {\bibfnamefont {L.}~\bibnamefont {Si}} \emph {et~al.} (\bibinfo {collaboration} {PandaX Collaboration}),\ }\href {\doibase 10.34133/2022/9798721} {\bibfield  {journal} {\bibinfo  {journal} {Research}\ }\textbf {\bibinfo {volume} {2022}},\ \bibinfo {pages} {9798721} (\bibinfo {year} {2022})}\BibitemShut {NoStop}%
\bibitem [{\citenamefont {Chen}\ \emph {et~al.}(2021)\citenamefont {Chen} \emph {et~al.}}]{BambooMC}%
  \BibitemOpen
  \bibfield  {author} {\bibinfo {author} {\bibfnamefont {X.}~\bibnamefont {Chen}} \emph {et~al.},\ }\href {\doibase 10.1088/1748-0221/16/09/T09004} {\bibfield  {journal} {\bibinfo  {journal} {J. Instrum.}\ }\textbf {\bibinfo {volume} {16}},\ \bibinfo {pages} {T09004} (\bibinfo {year} {2021})}\BibitemShut {NoStop}%
\bibitem [{\citenamefont {Agostinelli}\ \emph {et~al.}(2003)\citenamefont {Agostinelli} \emph {et~al.}}]{Geant4}%
  \BibitemOpen
  \bibfield  {author} {\bibinfo {author} {\bibfnamefont {S.}~\bibnamefont {Agostinelli}} \emph {et~al.},\ }\href {\doibase https://doi.org/10.1016/S0168-9002(03)01368-8} {\bibfield  {journal} {\bibinfo  {journal} {Nucl. Instrum. Methods Phys. Res. A}\ }\textbf {\bibinfo {volume} {506}},\ \bibinfo {pages} {250} (\bibinfo {year} {2003})}\BibitemShut {NoStop}%
\bibitem [{\citenamefont {Ma}\ \emph {et~al.}(2020)\citenamefont {Ma} \emph {et~al.}}]{P2_Rn220_Ma}%
  \BibitemOpen
  \bibfield  {author} {\bibinfo {author} {\bibfnamefont {W.}~\bibnamefont {Ma}} \emph {et~al.},\ }\href {\doibase 10.1088/1748-0221/15/12/P12038} {\bibfield  {journal} {\bibinfo  {journal} {J. Instrum.}\ }\textbf {\bibinfo {volume} {15}},\ \bibinfo {pages} {P12038} (\bibinfo {year} {2020})}\BibitemShut {NoStop}%
\bibitem [{\citenamefont {Yan}\ \emph {et~al.}(2024)\citenamefont {Yan} \emph {et~al.}}]{P4Xe134DBD}%
  \BibitemOpen
  \bibfield  {author} {\bibinfo {author} {\bibfnamefont {X.}~\bibnamefont {Yan}} \emph {et~al.} (\bibinfo {collaboration} {PandaX Collaboration}),\ }\href {\doibase 10.1103/PhysRevLett.132.152502} {\bibfield  {journal} {\bibinfo  {journal} {Phys. Rev. Lett.}\ }\textbf {\bibinfo {volume} {132}},\ \bibinfo {pages} {152502} (\bibinfo {year} {2024})}\BibitemShut {NoStop}%
\bibitem [{\citenamefont {Meng}\ \emph {et~al.}(2021)\citenamefont {Meng} \emph {et~al.}}]{P4CommissonRun}%
  \BibitemOpen
  \bibfield  {author} {\bibinfo {author} {\bibfnamefont {Y.}~\bibnamefont {Meng}} \emph {et~al.} (\bibinfo {collaboration} {PandaX Collaboration}),\ }\href {\doibase 10.1103/PhysRevLett.127.261802} {\bibfield  {journal} {\bibinfo  {journal} {Phys. Rev. Lett.}\ }\textbf {\bibinfo {volume} {127}},\ \bibinfo {pages} {261802} (\bibinfo {year} {2021})}\BibitemShut {NoStop}%
\bibitem [{\citenamefont {Aprile}\ \emph {et~al.}(2022{\natexlab{b}})\citenamefont {Aprile} \emph {et~al.}}]{XENON:Xe124}%
  \BibitemOpen
  \bibfield  {author} {\bibinfo {author} {\bibfnamefont {E.}~\bibnamefont {Aprile}} \emph {et~al.} (\bibinfo {collaboration} {XENON Collaboration}),\ }\href {\doibase 10.1103/PhysRevC.106.024328} {\bibfield  {journal} {\bibinfo  {journal} {Phys. Rev. C}\ }\textbf {\bibinfo {volume} {106}},\ \bibinfo {pages} {024328} (\bibinfo {year} {2022}{\natexlab{b}})}\BibitemShut {NoStop}%
\bibitem [{\citenamefont {Abdusalam}\ \emph {et~al.}(2022)\citenamefont {Abdusalam} \emph {et~al.}}]{Salam_AC}%
  \BibitemOpen
  \bibfield  {author} {\bibinfo {author} {\bibfnamefont {A.}~\bibnamefont {Abdusalam}} \emph {et~al.},\ }\href {http://iopscience.iop.org/article/10.1088/1674-1137/ac7cd8} {\bibfield  {journal} {\bibinfo  {journal} {Chinese Phys. C}\ } (\bibinfo {year} {2022})}\BibitemShut {NoStop}%
\bibitem [{\citenamefont {Baak}\ \emph {et~al.}(2015)\citenamefont {Baak}, \citenamefont {Besjes}, \citenamefont {Côté}, \citenamefont {Koutsman}, \citenamefont {Lorenz},\ and\ \citenamefont {Short}}]{baak_histfitter_2015}%
  \BibitemOpen
  \bibfield  {author} {\bibinfo {author} {\bibfnamefont {M.}~\bibnamefont {Baak}}, \bibinfo {author} {\bibfnamefont {G.~J.}\ \bibnamefont {Besjes}}, \bibinfo {author} {\bibfnamefont {D.}~\bibnamefont {Côté}}, \bibinfo {author} {\bibfnamefont {A.}~\bibnamefont {Koutsman}}, \bibinfo {author} {\bibfnamefont {J.}~\bibnamefont {Lorenz}}, \ and\ \bibinfo {author} {\bibfnamefont {D.}~\bibnamefont {Short}},\ }\href {\doibase https://doi.org/10.1140/epjc/s10052-015-3327-7} {\bibfield  {journal} {\bibinfo  {journal} {Eur. Phys. J. C}\ }\textbf {\bibinfo {volume} {75}},\ \bibinfo {pages} {153} (\bibinfo {year} {2015})}\BibitemShut {NoStop}%
\bibitem [{\citenamefont {Cranmer}\ \emph {et~al.}(2012)\citenamefont {Cranmer}, \citenamefont {Shibata}, \citenamefont {Verkerke}, \citenamefont {Moneta},\ and\ \citenamefont {Lewis}}]{Cranmer2012HistFactoryAT}%
  \BibitemOpen
  \bibfield  {author} {\bibinfo {author} {\bibfnamefont {K.}~\bibnamefont {Cranmer}}, \bibinfo {author} {\bibfnamefont {A.}~\bibnamefont {Shibata}}, \bibinfo {author} {\bibfnamefont {W.}~\bibnamefont {Verkerke}}, \bibinfo {author} {\bibfnamefont {L.}~\bibnamefont {Moneta}}, \ and\ \bibinfo {author} {\bibfnamefont {G.~H.}\ \bibnamefont {Lewis}},\ }\href {https://api.semanticscholar.org/CorpusID:115388506} {\bibfield  {journal} {\bibinfo  {journal} {CERNOPEN-2012-016}\ } (\bibinfo {year} {2012})}\BibitemShut {NoStop}%
\bibitem [{\citenamefont {O'Hare}(2020)}]{AxionLimits}%
  \BibitemOpen
  \bibfield  {author} {\bibinfo {author} {\bibfnamefont {C.}~\bibnamefont {O'Hare}},\ }\href {\doibase 10.5281/zenodo.3932430} {\enquote {\bibinfo {title} {cajohare/axionlimits: Axionlimits},}\ }\bibinfo {howpublished} {\url{https://cajohare.github.io/AxionLimits/}} (\bibinfo {year} {2020})\BibitemShut {NoStop}%
\bibitem [{\citenamefont {Akerib}\ \emph {et~al.}(2017)\citenamefont {Akerib} \emph {et~al.}}]{LUX_axion}%
  \BibitemOpen
  \bibfield  {author} {\bibinfo {author} {\bibfnamefont {D.~S.}\ \bibnamefont {Akerib}} \emph {et~al.} (\bibinfo {collaboration} {LUX Collaboration}),\ }\href {\doibase 10.1103/PhysRevLett.118.261301} {\bibfield  {journal} {\bibinfo  {journal} {Phys. Rev. Lett.}\ }\textbf {\bibinfo {volume} {118}},\ \bibinfo {pages} {261301} (\bibinfo {year} {2017})}\BibitemShut {NoStop}%
\bibitem [{\citenamefont {Viaux}\ \emph {et~al.}(2013)\citenamefont {Viaux}, \citenamefont {Catelan}, \citenamefont {Stetson}, \citenamefont {Raffelt}, \citenamefont {Redondo}, \citenamefont {Valcarce},\ and\ \citenamefont {Weiss}}]{RedGiant}%
  \BibitemOpen
  \bibfield  {author} {\bibinfo {author} {\bibfnamefont {N.}~\bibnamefont {Viaux}}, \bibinfo {author} {\bibfnamefont {M.}~\bibnamefont {Catelan}}, \bibinfo {author} {\bibfnamefont {P.~B.}\ \bibnamefont {Stetson}}, \bibinfo {author} {\bibfnamefont {G.~G.}\ \bibnamefont {Raffelt}}, \bibinfo {author} {\bibfnamefont {J.}~\bibnamefont {Redondo}}, \bibinfo {author} {\bibfnamefont {A.~A.~R.}\ \bibnamefont {Valcarce}}, \ and\ \bibinfo {author} {\bibfnamefont {A.}~\bibnamefont {Weiss}},\ }\href {\doibase 10.1103/PhysRevLett.111.231301} {\bibfield  {journal} {\bibinfo  {journal} {Phys. Rev. Lett.}\ }\textbf {\bibinfo {volume} {111}},\ \bibinfo {pages} {231301} (\bibinfo {year} {2013})}\BibitemShut {NoStop}%
\bibitem [{\citenamefont {Alessandria}\ \emph {et~al.}(2013)\citenamefont {Alessandria} \emph {et~al.}}]{Fe57_CUORE}%
  \BibitemOpen
  \bibfield  {author} {\bibinfo {author} {\bibfnamefont {F.}~\bibnamefont {Alessandria}} \emph {et~al.} (\bibinfo {collaboration} {CUORE collaboration}),\ }\href {\doibase 10.1088/1475-7516/2013/05/007} {\bibfield  {journal} {\bibinfo  {journal} {J. Cosmol. Astropart. Phys.}\ }\textbf {\bibinfo {volume} {2013}},\ \bibinfo {pages} {007} (\bibinfo {year} {2013})}\BibitemShut {NoStop}%
\bibitem [{\citenamefont {Wang}\ \emph {et~al.}(2020)\citenamefont {Wang} \emph {et~al.}}]{CDEX_axion_2019}%
  \BibitemOpen
  \bibfield  {author} {\bibinfo {author} {\bibfnamefont {Y.}~\bibnamefont {Wang}} \emph {et~al.} (\bibinfo {collaboration} {CDEX Collaboration}),\ }\href {\doibase 10.1103/PhysRevD.101.052003} {\bibfield  {journal} {\bibinfo  {journal} {Phys. Rev. D}\ }\textbf {\bibinfo {volume} {101}},\ \bibinfo {pages} {052003} (\bibinfo {year} {2020})}\BibitemShut {NoStop}%
\bibitem [{\citenamefont {Vinyoles}\ \emph {et~al.}(2015)\citenamefont {Vinyoles}, \citenamefont {Serenelli}, \citenamefont {Villante}, \citenamefont {Basu}, \citenamefont {Redondo},\ and\ \citenamefont {Isern}}]{SolarNu_axion}%
  \BibitemOpen
  \bibfield  {author} {\bibinfo {author} {\bibfnamefont {N.}~\bibnamefont {Vinyoles}}, \bibinfo {author} {\bibfnamefont {A.}~\bibnamefont {Serenelli}}, \bibinfo {author} {\bibfnamefont {F.}~\bibnamefont {Villante}}, \bibinfo {author} {\bibfnamefont {S.}~\bibnamefont {Basu}}, \bibinfo {author} {\bibfnamefont {J.}~\bibnamefont {Redondo}}, \ and\ \bibinfo {author} {\bibfnamefont {J.}~\bibnamefont {Isern}},\ }\href {\doibase 10.1088/1475-7516/2015/10/015} {\bibfield  {journal} {\bibinfo  {journal} {J. Cosmol. and Astropart. Phys.}\ }\textbf {\bibinfo {volume} {2015}},\ \bibinfo {pages} {015} (\bibinfo {year} {2015})}\BibitemShut {NoStop}%
\bibitem [{\citenamefont {Anastassopoulos}\ \emph {et~al.}(2017)\citenamefont {Anastassopoulos} \emph {et~al.}}]{CAST}%
  \BibitemOpen
  \bibfield  {author} {\bibinfo {author} {\bibfnamefont {V.}~\bibnamefont {Anastassopoulos}} \emph {et~al.} (\bibinfo {collaboration} {CAST Collaboration}),\ }\href {\doibase 10.1038/nphys4109} {\bibfield  {journal} {\bibinfo  {journal} {Nat. Phys.}\ }\textbf {\bibinfo {volume} {13}},\ \bibinfo {pages} {584} (\bibinfo {year} {2017})}\BibitemShut {NoStop}%
\bibitem [{\citenamefont {Córsico}\ \emph {et~al.}(2014)\citenamefont {Córsico}, \citenamefont {Althaus}, \citenamefont {Miller~Bertolami}, \citenamefont {Kepler},\ and\ \citenamefont {García-Berro}}]{Xenon1TwhiteDwarfmumag}%
  \BibitemOpen
  \bibfield  {author} {\bibinfo {author} {\bibfnamefont {A.~H.}\ \bibnamefont {Córsico}}, \bibinfo {author} {\bibfnamefont {L.~G.}\ \bibnamefont {Althaus}}, \bibinfo {author} {\bibfnamefont {M.~M.}\ \bibnamefont {Miller~Bertolami}}, \bibinfo {author} {\bibfnamefont {S.}~\bibnamefont {Kepler}}, \ and\ \bibinfo {author} {\bibfnamefont {E.}~\bibnamefont {García-Berro}},\ }\href {\doibase 10.1088/1475-7516/2014/08/054} {\bibfield  {journal} {\bibinfo  {journal} {JCAP}\ }\textbf {\bibinfo {volume} {08}},\ \bibinfo {pages} {054} (\bibinfo {year} {2014})},\ \Eprint {http://arxiv.org/abs/1406.6034} {arXiv:1406.6034 [astro-ph.SR]} \BibitemShut {NoStop}%
\bibitem [{\citenamefont {Miller~Bertolami}(2014)}]{WhiteDwarfmumag}%
  \BibitemOpen
  \bibfield  {author} {\bibinfo {author} {\bibfnamefont {M.~M.}\ \bibnamefont {Miller~Bertolami}},\ }\href {\doibase 10.1051/0004-6361/201322641} {\bibfield  {journal} {\bibinfo  {journal} {Astron. Astrophys.}\ }\textbf {\bibinfo {volume} {562}},\ \bibinfo {pages} {A123} (\bibinfo {year} {2014})}\BibitemShut {NoStop}%
\bibitem [{\citenamefont {Liao}\ \emph {et~al.}(2020)\citenamefont {Liao} \emph {et~al.}}]{HXMT}%
  \BibitemOpen
  \bibfield  {author} {\bibinfo {author} {\bibfnamefont {J.-Y.}\ \bibnamefont {Liao}} \emph {et~al.},\ }\href {\doibase https://doi.org/10.1016/j.jheap.2020.02.010} {\bibfield  {journal} {\bibinfo  {journal} {J. High Energy Astrophys.}\ }\textbf {\bibinfo {volume} {27}},\ \bibinfo {pages} {24} (\bibinfo {year} {2020})}\BibitemShut {NoStop}%
\bibitem [{\citenamefont {Ng}\ \emph {et~al.}(2019)\citenamefont {Ng}, \citenamefont {Roach}, \citenamefont {Perez}, \citenamefont {Beacom}, \citenamefont {Horiuchi}, \citenamefont {Krivonos},\ and\ \citenamefont {Wik}}]{NuStarM31}%
  \BibitemOpen
  \bibfield  {author} {\bibinfo {author} {\bibfnamefont {K.~C.~Y.}\ \bibnamefont {Ng}}, \bibinfo {author} {\bibfnamefont {B.~M.}\ \bibnamefont {Roach}}, \bibinfo {author} {\bibfnamefont {K.}~\bibnamefont {Perez}}, \bibinfo {author} {\bibfnamefont {J.~F.}\ \bibnamefont {Beacom}}, \bibinfo {author} {\bibfnamefont {S.}~\bibnamefont {Horiuchi}}, \bibinfo {author} {\bibfnamefont {R.}~\bibnamefont {Krivonos}}, \ and\ \bibinfo {author} {\bibfnamefont {D.~R.}\ \bibnamefont {Wik}},\ }\href {\doibase 10.1103/PhysRevD.99.083005} {\bibfield  {journal} {\bibinfo  {journal} {Phys. Rev. D}\ }\textbf {\bibinfo {volume} {99}},\ \bibinfo {pages} {083005} (\bibinfo {year} {2019})}\BibitemShut {NoStop}%
\bibitem [{\citenamefont {Bouchet}\ \emph {et~al.}(2011)\citenamefont {Bouchet}, \citenamefont {Strong}, \citenamefont {Porter}, \citenamefont {Moskalenko}, \citenamefont {Jourdain},\ and\ \citenamefont {Roques}}]{INTEGRAL_Bouchet_2011}%
  \BibitemOpen
  \bibfield  {author} {\bibinfo {author} {\bibfnamefont {L.}~\bibnamefont {Bouchet}}, \bibinfo {author} {\bibfnamefont {A.~W.}\ \bibnamefont {Strong}}, \bibinfo {author} {\bibfnamefont {T.~A.}\ \bibnamefont {Porter}}, \bibinfo {author} {\bibfnamefont {I.~V.}\ \bibnamefont {Moskalenko}}, \bibinfo {author} {\bibfnamefont {E.}~\bibnamefont {Jourdain}}, \ and\ \bibinfo {author} {\bibfnamefont {J.-P.}\ \bibnamefont {Roques}},\ }\href {\doibase 10.1088/0004-637X/739/1/29} {\bibfield  {journal} {\bibinfo  {journal} {The Astrophysical Journal}\ }\textbf {\bibinfo {volume} {739}},\ \bibinfo {pages} {29} (\bibinfo {year} {2011})}\BibitemShut {NoStop}%
\bibitem [{\citenamefont {Aprile}\ \emph {et~al.}(2020)\citenamefont {Aprile} \emph {et~al.}}]{XENON1T_excess}%
  \BibitemOpen
  \bibfield  {author} {\bibinfo {author} {\bibfnamefont {E.}~\bibnamefont {Aprile}} \emph {et~al.} (\bibinfo {collaboration} {XENON Collaboration}),\ }\href {\doibase 10.1103/PhysRevD.102.072004} {\bibfield  {journal} {\bibinfo  {journal} {Phys. Rev. D}\ }\textbf {\bibinfo {volume} {102}},\ \bibinfo {pages} {072004} (\bibinfo {year} {2020})}\BibitemShut {NoStop}%
\bibitem [{\citenamefont {Kim}(1979)}]{KSVZ_1}%
  \BibitemOpen
  \bibfield  {author} {\bibinfo {author} {\bibfnamefont {J.~E.}\ \bibnamefont {Kim}},\ }\href {\doibase 10.1103/PhysRevLett.43.103} {\bibfield  {journal} {\bibinfo  {journal} {Phys. Rev. Lett.}\ }\textbf {\bibinfo {volume} {43}},\ \bibinfo {pages} {103} (\bibinfo {year} {1979})}\BibitemShut {NoStop}%
\bibitem [{\citenamefont {Shifman}\ \emph {et~al.}(1980)\citenamefont {Shifman}, \citenamefont {Vainshtein},\ and\ \citenamefont {Zakharov}}]{KSVZ_2}%
  \BibitemOpen
  \bibfield  {author} {\bibinfo {author} {\bibfnamefont {M.}~\bibnamefont {Shifman}}, \bibinfo {author} {\bibfnamefont {A.}~\bibnamefont {Vainshtein}}, \ and\ \bibinfo {author} {\bibfnamefont {V.}~\bibnamefont {Zakharov}},\ }\href {\doibase https://doi.org/10.1016/0550-3213(80)90209-6} {\bibfield  {journal} {\bibinfo  {journal} {Nuclear Physics B}\ }\textbf {\bibinfo {volume} {166}},\ \bibinfo {pages} {493} (\bibinfo {year} {1980})}\BibitemShut {NoStop}%
\bibitem [{\citenamefont {Dine}\ \emph {et~al.}(1981)\citenamefont {Dine}, \citenamefont {Fischler},\ and\ \citenamefont {Srednicki}}]{DFSZ_1}%
  \BibitemOpen
  \bibfield  {author} {\bibinfo {author} {\bibfnamefont {M.}~\bibnamefont {Dine}}, \bibinfo {author} {\bibfnamefont {W.}~\bibnamefont {Fischler}}, \ and\ \bibinfo {author} {\bibfnamefont {M.}~\bibnamefont {Srednicki}},\ }\href {\doibase https://doi.org/10.1016/0370-2693(81)90590-6} {\bibfield  {journal} {\bibinfo  {journal} {Phys. Lett. B}\ }\textbf {\bibinfo {volume} {104}},\ \bibinfo {pages} {199} (\bibinfo {year} {1981})}\BibitemShut {NoStop}%
\bibitem [{\citenamefont {Zhitnitskii}(1980)}]{DFSZ_2}%
  \BibitemOpen
  \bibfield  {author} {\bibinfo {author} {\bibfnamefont {A.~P.}\ \bibnamefont {Zhitnitskii}},\ }\href {https://www.osti.gov/biblio/7063072} {\bibfield  {journal} {\bibinfo  {journal} {Sov. J. Nucl. Phys. (Engl. Transl.); (United States)}\ }\textbf {\bibinfo {volume} {31:2}} (\bibinfo {year} {1980})}\BibitemShut {NoStop}%
\bibitem [{\citenamefont {Cowan}\ \emph {et~al.}(2011{\natexlab{a}})\citenamefont {Cowan}, \citenamefont {Cranmer}, \citenamefont {Gross},\ and\ \citenamefont {Vitells}}]{Cowan:2010js}%
  \BibitemOpen
  \bibfield  {author} {\bibinfo {author} {\bibfnamefont {G.}~\bibnamefont {Cowan}}, \bibinfo {author} {\bibfnamefont {K.}~\bibnamefont {Cranmer}}, \bibinfo {author} {\bibfnamefont {E.}~\bibnamefont {Gross}}, \ and\ \bibinfo {author} {\bibfnamefont {O.}~\bibnamefont {Vitells}},\ }\href {\doibase 10.1140/epjc/s10052-011-1554-0} {\bibfield  {journal} {\bibinfo  {journal} {Eur. Phys. J. C}\ }\textbf {\bibinfo {volume} {71}},\ \bibinfo {pages} {1554} (\bibinfo {year} {2011}{\natexlab{a}})},\ \bibinfo {note} {[Erratum: Eur.Phys.J.C 73, 2501 (2013)]},\ \Eprint {http://arxiv.org/abs/1007.1727} {arXiv:1007.1727 [physics.data-an]} \BibitemShut {NoStop}%
\bibitem [{\citenamefont {Baxter}\ \emph {et~al.}(2021)\citenamefont {Baxter}, \citenamefont {Bloch}, \citenamefont {Bodnia} \emph {et~al.}}]{Baxter:PLR_conventions}%
  \BibitemOpen
  \bibfield  {author} {\bibinfo {author} {\bibfnamefont {D.}~\bibnamefont {Baxter}}, \bibinfo {author} {\bibfnamefont {I.~M.}\ \bibnamefont {Bloch}}, \bibinfo {author} {\bibfnamefont {E.}~\bibnamefont {Bodnia}},  \emph {et~al.},\ }\href {\doibase 10.1140/epjc/s10052-021-09655-y} {\bibfield  {journal} {\bibinfo  {journal} {Eur. Phys. J. C}\ }\textbf {\bibinfo {volume} {81}},\ \bibinfo {pages} {907} (\bibinfo {year} {2021})}\BibitemShut {NoStop}%
\bibitem [{\citenamefont {Cowan}\ \emph {et~al.}(2011{\natexlab{b}})\citenamefont {Cowan}, \citenamefont {Cranmer}, \citenamefont {Gross},\ and\ \citenamefont {Vitells}}]{cowan2011powerconstrainedlimits}%
  \BibitemOpen
  \bibfield  {author} {\bibinfo {author} {\bibfnamefont {G.}~\bibnamefont {Cowan}}, \bibinfo {author} {\bibfnamefont {K.}~\bibnamefont {Cranmer}}, \bibinfo {author} {\bibfnamefont {E.}~\bibnamefont {Gross}}, \ and\ \bibinfo {author} {\bibfnamefont {O.}~\bibnamefont {Vitells}},\ }\href {https://arxiv.org/abs/1105.3166} {\  (\bibinfo {year} {2011}{\natexlab{b}})},\ \Eprint {http://arxiv.org/abs/1105.3166} {arXiv:1105.3166 [physics.data-an]} \BibitemShut {NoStop}%
\end{thebibliography}%

\end{document}